\documentclass{aa} 
\usepackage{graphicx}
\usepackage{comment}
\usepackage{txfonts}
\usepackage{xcolor}
\usepackage{mathrsfs}
\usepackage{amssymb,amsmath}
\usepackage{newtxtext,newtxmath}
\usepackage{amsmath}	
\usepackage{amssymb}	
\DeclareMathAlphabet{\mathbi}{OT1}{ptm}{bx}{it}
\SetMathAlphabet\mathbi{bold}{OT1}{ptm}{bx}{it}

\usepackage{hyperref}
\hypersetup{
    colorlinks=true,
    citecolor=blue,
    linkcolor=blue,
    filecolor=magenta,      
    urlcolor=cyan,
}
\begin{document}
   \title{Accretion-disk sizes in two quasars with interferometrically resolved broad-line regions at $z=2.3$ and $z=4.0$}
   \titlerunning{Accretion-disk sizes in quasars with resolved BLRs}
    \author{F. Pozo Nu\~nez \inst{1}\href{https://orcid.org/0000-0002-6716-4179}{\includegraphics[scale=0.5]{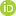}}
          \and
          E. Ba\~nados \inst{2}\href{https://orcid.org/0000-0002-2931-7824}{\includegraphics[scale=0.5]{orcidicon.png}}
          \and
          S. Panda\inst{3}\thanks{Gemini Science Fellow}\href{https://orcid.org/0000-0002-5854-7426}{\includegraphics[scale=0.5]{orcidicon.png}}
          \and
          A. K. Mandal\inst{4}\href{https://orcid.org/0000-0001-9957-6349}{\includegraphics[scale=0.5]{orcidicon.png}}
          \and
          B. Czerny\inst{4}\href{https://orcid.org/0000-0001-5848-4333}{\includegraphics[scale=0.5]{orcidicon.png}}
          \and
          J. Heidt\inst{5}\href{https://orcid.org/0000-0002-0320-1292}{\includegraphics[scale=0.5]{orcidicon.png}}
          }

   \institute{Astroinformatics, Heidelberg Institute for Theoretical Studies, Schloss-Wolfsbrunnenweg 35, 69118 Heidelberg, Germany\\
   \email{francisco.pozon@gmail.com}
   \and
   Max-Planck Institut f\"ur Astronomie, K{\"o}nigstuhl 17 Heidelberg, Germany
   \and
   International Gemini Observatory/NSF NOIRLab, Casilla 603, La Serena, Chile
   \and
   Center for Theoretical Physics, Polish Academy of Sciences, Al. Lotnik\'ow 32/46, 02-668 Warsaw, Poland 
   \and
   Landessternwarte, Zentrum f\"ur Astronomie der Universit\"at Heidelberg, K{\"o}nigstuhl 12, 69117 Heidelberg, Germany \\
              }
   \date{21 June, 2026; accepted 28 August, 2026}

\abstract{
We present the first direct comparison between accretion-disk (AD) and 
broad-line region (BLR) sizes in quasars at $z > 2$, enabled by combining 
continuum reverberation mapping with interferometric BLR constraints from 
GRAVITY and GRAVITY+. Using medium-band photometric monitoring with the 
MPG/ESO 2.2\,m telescope, we measure inter-band continuum lags in two  quasars: SDSS~J092034.17+065718.0 at $z = 2.33$ (J0920) and 
SMSS~J052915.80$-$435152.0 at $z = 3.96$ (J0529), among the most luminous quasars 
known ($L_{\rm bol} \sim 10^{48}$\,erg\,s$^{-1}$). These are currently the 
only two quasars at $z > 2$ with spatially resolved BLRs and dynamical 
black-hole masses from interferometry, making this comparison uniquely 
possible. We detect significant continuum lags in both quasars, increasing 
monotonically with wavelength. The inferred UV disk sizes are 
$R_{\rm AD} = 4.35^{+0.78}_{-0.91}$\,light-days for J0529 and 
$R_{\rm AD} = 3.15^{+0.50}_{-0.48}$\,light-days for J0920. For J0920, 
accreting at $\lambda_{\rm Edd} \sim 7$--20, the disk size is consistent with standard thin-disk expectations, a 
surprising result given that the standard thin-disk approximation is not 
expected to hold in this super-Eddington regime. For J0529, the disk size 
is consistent with thin-disk predictions under the single-epoch black-hole 
mass, but would imply disk inflation by a 
factor of a few if the GRAVITY+ dynamical mass is adopted instead --- an 
order of magnitude lower. This tension illustrates that UV continuum disk 
sizes provide an independent physical scale that any self-consistent model 
of the black-hole mass and accretion rate must satisfy, adding leverage in 
objects where BLR kinematics are dominated by outflows rather than virial 
motion. A comparison with the interferometric BLR sizes reveals pronounced 
radial hierarchies, with $R_{\rm BLR}/R_{\rm AD} \sim 270$ for J0529 (H$\beta$) and 
$\sim\!115$ for J0920 (H$\alpha$). The successful detection of lags in both quasars at $L_{\rm bol} \sim 
10^{48}$\,erg\,s$^{-1}$ demonstrates that continuum reverberation mapping 
remains feasible even at the low-amplitude variability end expected for the 
most luminous systems, opening a path toward extending accretion-disk size 
measurements to larger samples with wide-field surveys such as the Vera 
C.\ Rubin Observatory's LSST. 
}


   \maketitle

\section{Introduction}

Accretion onto supermassive black holes (SMBHs) powers active galactic nuclei (AGN) and determines their radiative output across cosmic time. 
In the framework of the standard geometrically thin, optically thick accretion disk (\citealt{Shakura1973,Novikov1973}), the disk temperature decreases with radius as $T(R)\propto R^{-3/4}$, implying a characteristic size--wavelength relation $R \propto \lambda^{4/3}$. 
Continuum reverberation mapping (CRM) provides a direct observational test of this picture by measuring inter-band time delays that trace light-travel times across the disk \citep[e.g.,][]{Collier1998,Sergeev2005,Cackett2007}. 
In this scenario, variability originating in the innermost disk propagates outwards, such that longer-wavelength emission responds with measurable delays relative to the ultraviolet (UV) continuum.

Over the past decade, multiwavelength monitoring campaigns have produced increasingly precise continuum-lag measurements in nearby Seyferts and moderate-luminosity AGN \citep[e.g.,][]{McHardy2014,Fausnaugh2016,Cackett2018,Edelson2019,HernandezSantisteban2020}, as well as larger reverberation mapping samples such as SDSS-RM \citep[e.g.,][]{Homayouni2019,Sharp2024}.
A recurring result is that observed continuum-emitting regions are often larger than predicted by the simplest thin-disk model, typically by factors of $\sim2$--$3$ \citep[e.g.,][]{Cackett2018,Gonzales2023,Thorne25,Mandal25}.
The origin of this apparent ``disk inflation'' remains debated.
Proposed explanations include contamination by diffuse continuum emission from the broad-line region (BLR) \citep{Korista2019,Chelouche2019}, biases in the inferred luminosity due to internal reddening (\citealt{Gaskell2017,Gaskell2023}), inhomogeneous disk structures \citep{Dexter2011,Hall2018}, reprocessing geometries involving elevated X-ray coronae increasing light-travel delays \citep{Kammoun2023}, systematic shifts in thin-disk predictions arising from inclination-dependent black hole mass estimates \citep{Pozo2019}, and deviations from the canonical temperature profile \citep{Weaver2022,Papadakis2022}.
Disentangling these effects therefore requires continuum reverberation measurements spanning a wider range of luminosity, redshift, and accretion state.

To date, most CRM measurements and accretion disk size constraints have been limited to AGN at $z\leq 1$, with a few notable higher-redshift exceptions provided by gravitationally lensed quasars at $ 1 \leq z\leq 2.5$ (\citealt{Bate2008,Cornachione2020,Marculewicz2024,Sorgenfrei2025}). 
An even more distant case has recently been reported for the lensed quasar J043947.08+163415.7 at $z=6.5$, where \citet{Leung2026} identified variability and tentative accretion-disk signatures broadly consistent with standard thin-disk theory.
Extending such studies to unlensed luminous quasars at $z \gtrsim 2$ is therefore a key step toward testing accretion-disk structure under more extreme conditions near the peak epoch of black hole growth at $z\sim$2-4.
In this regime, black holes can accrete at high or even super-Eddington rates; here, the standard thin-disk approximation is expected to break down, and alternative accretion-flow geometries may become relevant \citep{Abramowicz1988,Sadowski2009}. 
Although Super-Eddington accretion has been explored observationally in lower-redshift reverberation-mapped AGN \citep[e.g.,][]{Du2015,Bai2026},
and accretion-disk continuum reverberation has been measured in the
super-Eddington source Mrk\,142 \citep{Cackett2020}, direct continuum lag measurements for high-redshift quasars remain hampered by cosmological time dilation, the need for long observing baselines, and high requirements for photometric precision.
Broadening these measurements to larger samples of high-luminosity systems is therefore essential to determine how accretion-disk structure scales with luminosity, accretion regime, and cosmic epoch (\citealt{Pozo2025,Steyn26}).

A particularly powerful opportunity arises when CRM measurements can be combined with independent constraints on the BLR. 
Recent interferometric observations with GRAVITY and GRAVITY+ have spatially resolved BLRs in a growing sample of $\sim 10$ AGN and quasars, enabling direct measurements of their sizes and dynamical black-hole masses \citep{Gravity2018,Gravity2024}. 
This opens the possibility of comparing the radial scale of the UV continuum-emitting disk with that of the BLR in the same objects, and to test whether the inferred disk size is consistent with independently constrained global source parameters.

In this work we focus on two luminous quasars that are, to date, the only objects at $z>2$ for which interferometric observations have spatially resolved the BLR and yielded a dynamical black hole mass.
This makes the disk-to-BLR size comparison presented here uniquely possible.
The first is SDSS J092034.17+065718.0 (hereafter J0920), a luminous quasar at $z=2.325$ for which GRAVITY resolved the H$\alpha$ BLR and measured a dynamical black-hole mass \citep{Abuter2024}.
The second is SMSS J052915.80-435152.0 (hereafter J0529), one of the most luminous quasars known, at $z=3.962$ (\citealt{Wolf2024}), for which GRAVITY+ resolved the H$\beta$ BLR, constrained its outflow-dominated geometry, and measured a dynamical black-hole mass \citep{Gravity2026}.
Together, these two objects probe a regime of luminosity, accretion state, and black-hole growth that remains largely unexplored by continuum reverberation studies.

We present medium-band photometric monitoring of J0920 and J0529 obtained with the MPG/ESO 2.2 m telescope. 
The selected filters sample continuum-dominated rest-frame UV emission, allowing us to measure inter-band continuum lags and infer characteristic accretion-disk scales. 
Our approach builds on the observing strategy and analysis framework of \cite{Pozo2025}, who demonstrated the feasibility of measuring the accretion disk size of a $z\sim 2$ quasar from ground-based medium-band monitoring, but without the benefit of independent interferometric BLR constraints.
We then compare the measured lag spectra with nominal thin-disk expectations and place the inferred disk sizes in the context of the interferometrically measured BLR scales.

The paper is organized as follows. Section~2 describes the observations and data reduction. Section~3 presents the light curves, lag measurements, and inferred disk sizes.
Section~4 discusses the implications for thin-disk interpretations and for the inner structure of luminous high-$z$ quasars. 
Section 5 summarizes our conclusions.
Throughout this work, we assume a concordance cosmology with ${H_{0}=70\ \mathrm{km\ s^{-1}\ Mpc^{-1}}}$, $\Omega_{\Lambda}=0.73$ and $\Omega_{m}=0.27$.

\begin{figure}
\centering
\includegraphics[width=\columnwidth]{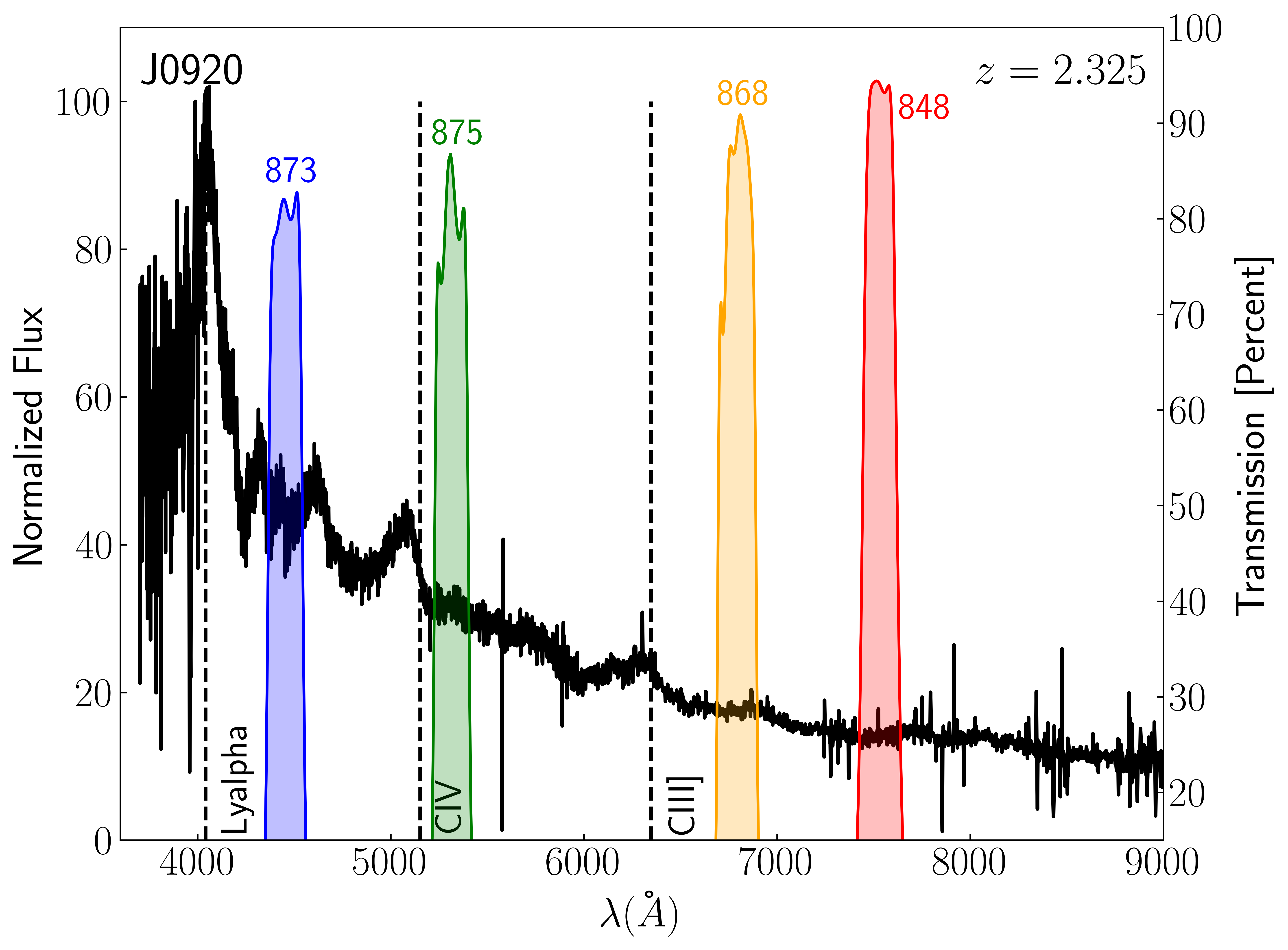}
\includegraphics[width=\columnwidth]{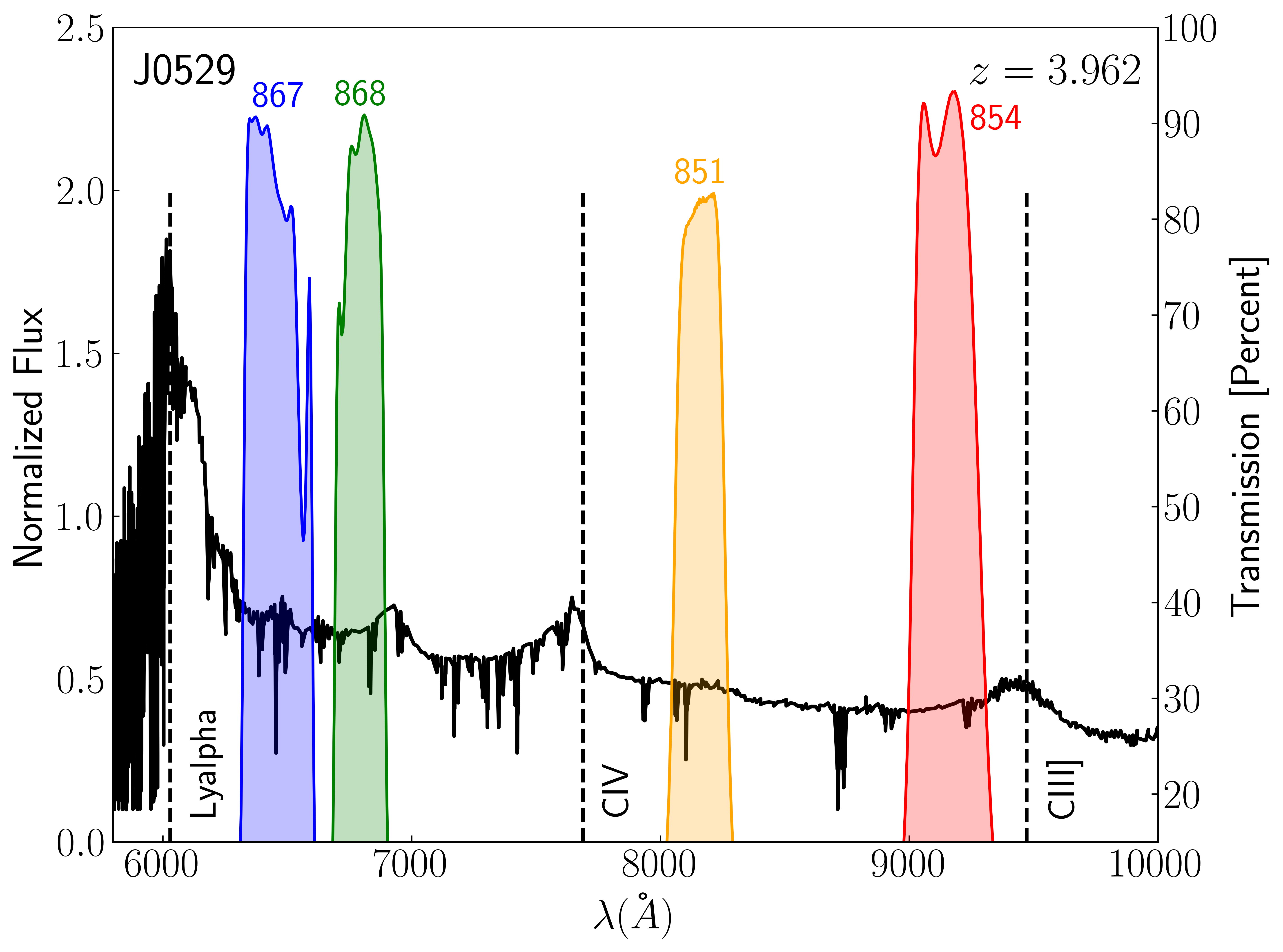}
\caption{
Observed-frame spectra of J0920 (top) and J0529 (bottom) with the WFI medium-band filter transmission windows overlaid.
The dashed vertical lines mark the positions of prominent broad emission lines in each spectrum.
The spectrum of J0920 is taken from the LAMOST survey (\citealt{Zhao2012}), while the spectrum of J0529 is reproduced from \citet{Wolf2024}. 
The filters were selected to probe continuum-dominated regions while avoiding the strongest broad emission features.
}
\label{fig:spectra_filters}
\end{figure}

\begin{table*}
\caption{Physical parameters of the target quasars adopted from the literature.}
\label{tab:physical_parameters}
\centering
\begin{tabular}{lcccccc}
\hline\hline
Object & Redshift & $M_{\rm BH}^{\rm dyn}$ & $M_{\rm BH}^{\rm SE}$ & $L_{\rm bol}$ & $\lambda_{\rm Edd}$ & $\dot{M}$ \\
 & $z$ & ($M_\odot$) & ($M_\odot$) & (erg s$^{-1}$) & & ($M_\odot$ yr$^{-1}$) \\
(1) & (2) & (3) & (4) & (5) & (6) & (7) \\
\hline
SDSS\,J092034.17$+$065718.0 & 2.325
& $(3.2^{+2.8}_{-1.5})\times10^{8}$
& $(0.9$--$5)\times10^{9}$
& $(1.6$--$8.0)\times10^{47}$
& $\sim7$--$20$
& $\sim28$--$141$ \\

SMSS\,J052915.80$-$435152.0 & 3.962
& $(7.9^{+2.3}_{-2.1})\times10^{8}$
& $(1.9\pm0.6)\times10^{10}$
& $1.9\times10^{48}$
& $\sim6$--$19$
& $\sim335$ \\
\hline
\end{tabular}
\tablefoot{
Columns: (1) object; (2) redshift; (3) dynamical black-hole mass from GRAVITY/GRAVITY+; (4) single-epoch virial black-hole mass; (5) bolometric luminosity; (6) Eddington ratio computed from the dynamical mass; (7) mass accretion rate.
The dynamical mass of J0920 is from \citet{Abuter2024}, obtained from the 
spatially resolved H$\alpha$ BLR; its single-epoch range spans the H$\alpha$ ($\sim0.9\times10^{9}$) to C\,IV ($\sim5\times10^{9}$) estimates reported therein, the latter biased high by a strong non-virial C\,IV component.
For J0529, the dynamical GRAVITY+ mass is from \citet{Gravity2026}, derived from the spatially resolved H$\beta$ BLR, while the single-epoch mass is the all-method best estimate of \citet{Wolf2024}.
The Eddington ratio and accretion rate assume a radiative efficiency $\eta=0.1$, with $\dot{M}=L_{\rm bol}/(\eta c^{2})$; using the single-epoch masses instead gives $\lambda_{\rm Edd}\sim0.2$--$1$ for J0920 and $\sim0.9$ for J0529; for J0920 this lower value reflects the 
overestimated single-epoch mass (Sect.~\ref{res:disksize}) rather than a genuinely lower accretion rate.
}
\end{table*}

\begin{table*}
\caption{Summary of the WFI medium-band filters used for the photometric monitoring with the MPG/ESO 2.2\,m telescope at La Silla.}
\label{tab:filters}
\centering
\begin{tabular}{lcccc}
\hline\hline
Object & Filter & $\lambda$ range (\AA) & CWL (\AA) & FWHM (\AA) \\
\hline
SMSS\,J052915.80$-$435152.0 & 867 & 6325--6601 & 6463 & 277 \\
      & 868 & 6695--6891 & 6793 & 197 \\
      & 851 & 8054--8264 & 8159 & 209 \\
      & 854 & 9010--9286 & 9148 & 275 \\
\hline
SDSS\,J092034.17$+$065718.0 & 873 & 4364--4546 & 4455 & 182 \\
      & 875 & 5227--5403 & 5315 & 177 \\
      & 868 & 6695--6891 & 6793 & 197 \\
      & 848 & 7441--7623 & 7532 & 183 \\
\hline
\end{tabular}
\tablefoot{
CWL denotes the central wavelength of the filter, and FWHM the full width at half maximum. The wavelength range is ${\rm CWL} \pm {\rm FWHM}/2$.
}
\end{table*}

\section{Observations and data reduction}\label{sec2}

Photometric monitoring was carried out with the MPG/ESO 2.2\,m telescope at La Silla Observatory using the Wide Field Imager (WFI) between October 2024 and April 2025 as part of an ongoing RM campaign targeting high-$z$ quasars. 
The WFI camera provides a field of view of $34'\times33'$, sampled by eight CCD detectors with a pixel scale of $0.238''$ pixel$^{-1}$ (\citealt{Baade1999}).
Medium-band filters were selected to sample continuum-dominated regions while minimizing contamination from strong broad emission lines. 
They provide an effective compromise between broad-band and narrow-band photometry, helping to isolate the continuum while retaining sufficient throughput for efficient monitoring.
A detailed description of the telescope, instrument setup, and data reduction procedures is given in the Appendix of \citet{Pozo2025}.
In brief, the raw images were processed following standard optical image-reduction procedures, including bias subtraction, flat-field correction, astrometric calibration, and correction for geometric distortions.
AGN light curves were extracted using aperture photometry relative to several non-variable reference stars located within the same WFI field of view. 
The aperture size was optimized to maximize the signal-to-noise ratio while minimizing contamination from the host galaxy and sky background. 
We adopted a circular aperture radius of $1.5''$, which yields stable photometric measurements with minimal flux scatter. 
Absolute flux calibration was performed using high-quality calibration stars from the DES DR2 catalog (\citealt{2021ApJS..255...20A}) located in the same field. 
The measured fluxes were corrected for atmospheric extinction at La Silla Observatory and for Galactic foreground extinction using the dust maps of \citet{Schlafly2011}.

J0529 was monitored in four medium bands centered at observed-frame wavelengths of 6463\,\AA, 6793\,\AA, 8159\,\AA, and 9148\,\AA, corresponding to WFI filter IDs 867, 868, 851, and 854, respectively.
J0920 was observed in four medium bands centered at 4455\,\AA, 5315\,\AA, 6793\,\AA, and 7532\,\AA, corresponding to WFI filter IDs 873, 875, 868, and 848, respectively.
The placement of the WFI filters relative to the observed spectra is illustrated in Fig.~\ref{fig:spectra_filters}.
For J0920, we show the archival LAMOST optical spectrum (\citealt{Abuter2024}), whereas for J0529 we adopt the optical spectrum presented in the discovery paper by \cite{Wolf2024}. 
The physical properties of J0920 and J0529 are summarized in Table~\ref{tab:physical_parameters}, while the adopted filter characteristics are listed in Table~\ref{tab:filters}.

Both quasars have single-epoch virial and spectro-interferometric
(dynamical) black-hole mass estimates, and in both cases the dynamical value is
the smaller of the two. Because the predicted thin-disk scale depends on the
black-hole mass as $A_{\rm pred}\propto(M_{\rm BH}\dot{M})^{1/3}$
(Sect.~\ref{normalcompar}), we briefly summarize these estimates and state the
values adopted throughout.
For J0920, \citet{Abuter2024} resolve the H$\alpha$ BLR with GRAVITY and find
that the line-emitting gas is consistent with a moderately inclined rotating
disk, yielding $\log(M_{\rm BH}/M_\odot)=8.51^{+0.27}_{-0.28}$. The
single-epoch estimates reported in the same work span more than an order of
magnitude, from $\log(M_{\rm BH}/M_\odot)=8.94\pm0.48$ (H$\alpha$) and
$9.24\pm0.47$ (H$\beta$) to $\approx9.7$ (C\,IV), the latter biased high by a
strongly blueshifted, non-virial component; an Eddington-ratio-corrected
H$\beta$ mass ($\approx8.6$\,dex) agrees with the dynamical value to within
$0.1$\,dex. Because the BLR kinematics are rotation-dominated and the dynamical
mass requires no virial scaling factor, we adopt it as the fiducial value for
J0920.

For J0529 the situation is less clear-cut. \citet{Gravity2026} resolve
the H$\beta$ BLR with GRAVITY+ and obtain
$\log(M_{\rm BH}/M_\odot)=8.90^{+0.11}_{-0.13}$, but find that the
line-emitting gas is dominated by a strong outflow contributing $\sim83\%$ of
the total line flux, so that the inferred mass depends more
sensitively on the assumed kinematic model than in the rotation-dominated case of J0920. It also lies
$\sim1.4$\,dex below the single-epoch value of
$(1.9\pm0.6)\times10^{10}\,M_\odot$ obtained by \citet{Wolf2024} from the
combined C\,IV, Mg\,II, and continuum-shape estimates. Given the
outflow-dominated BLR and the size of this discrepancy, we adopt the
single-epoch mass as the fiducial value for J0529, while regarding the
dynamical mass as an equally plausible alternative. Results are quoted for both mass estimates in each object.

\begin{figure*}
\centering
\includegraphics[width=\textwidth]{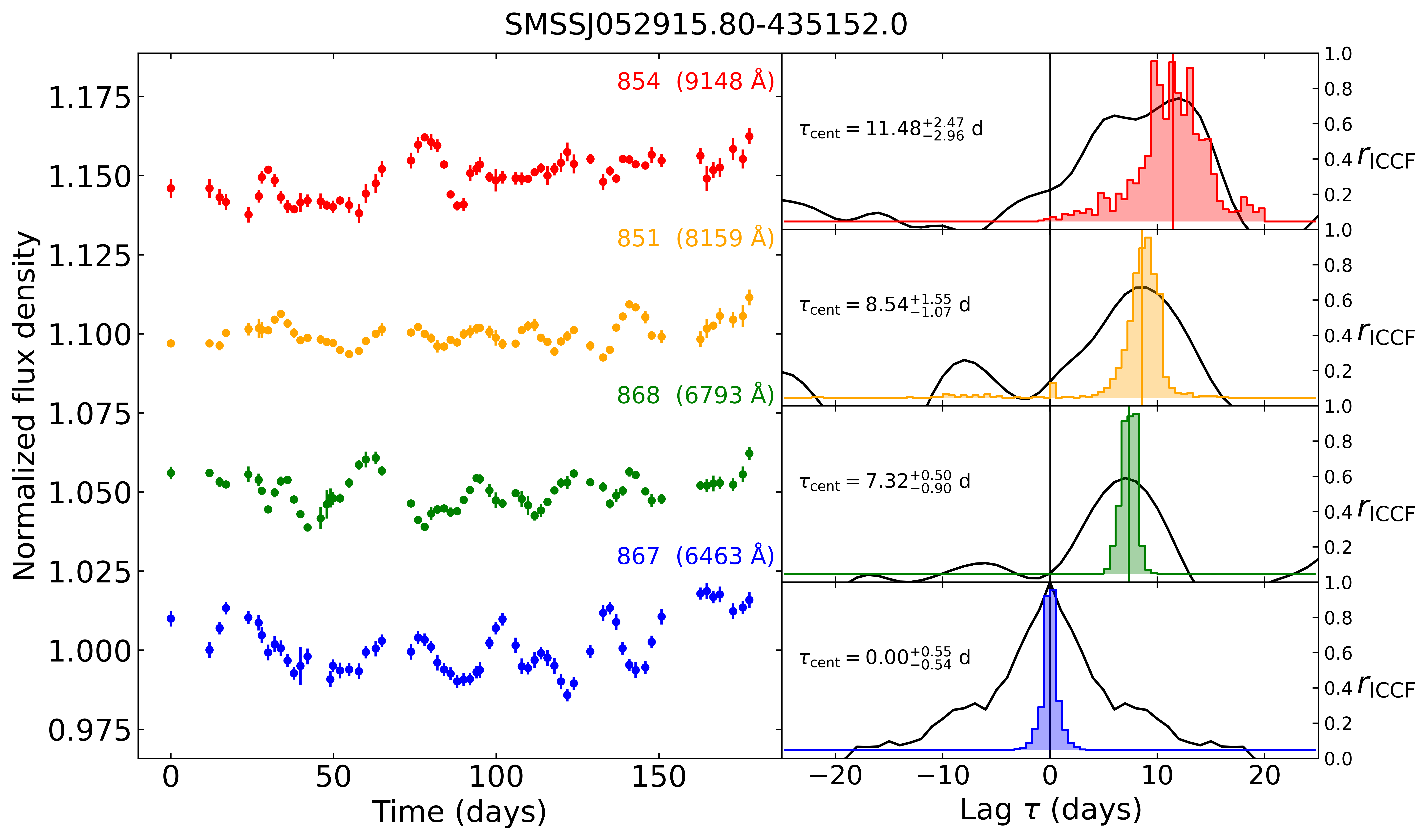}
\caption{Continuum light curves and cross-correlation analysis for J0529. 
\textit{Left panel:} Normalized light curves obtained with the WFI medium-band filters 867 (reference), 868, 851, and 854. 
The labels indicate the observed-frame central wavelength of each filter. 
The 867 band ($\lambda_{\rm obs}=6463$\,\AA) is used as the reference continuum band and is shown at the bottom, while the other light curves are vertically offset for clarity.
\textit{Right panels:} Interpolated cross-correlation functions (ICCF; black curves) between each band and the reference light curve. 
The histograms show the centroid lag distributions obtained from the flux randomization and random subset sampling (FR/RSS) Monte Carlo procedure. 
The vertical black line marks the zero-lag reference.
Median centroid lags from FR/RSS distributions are marked in each panel.}
\label{fig:lcs_lags_SMSS}
\end{figure*}

\begin{figure*}
\centering
\includegraphics[width=\textwidth]{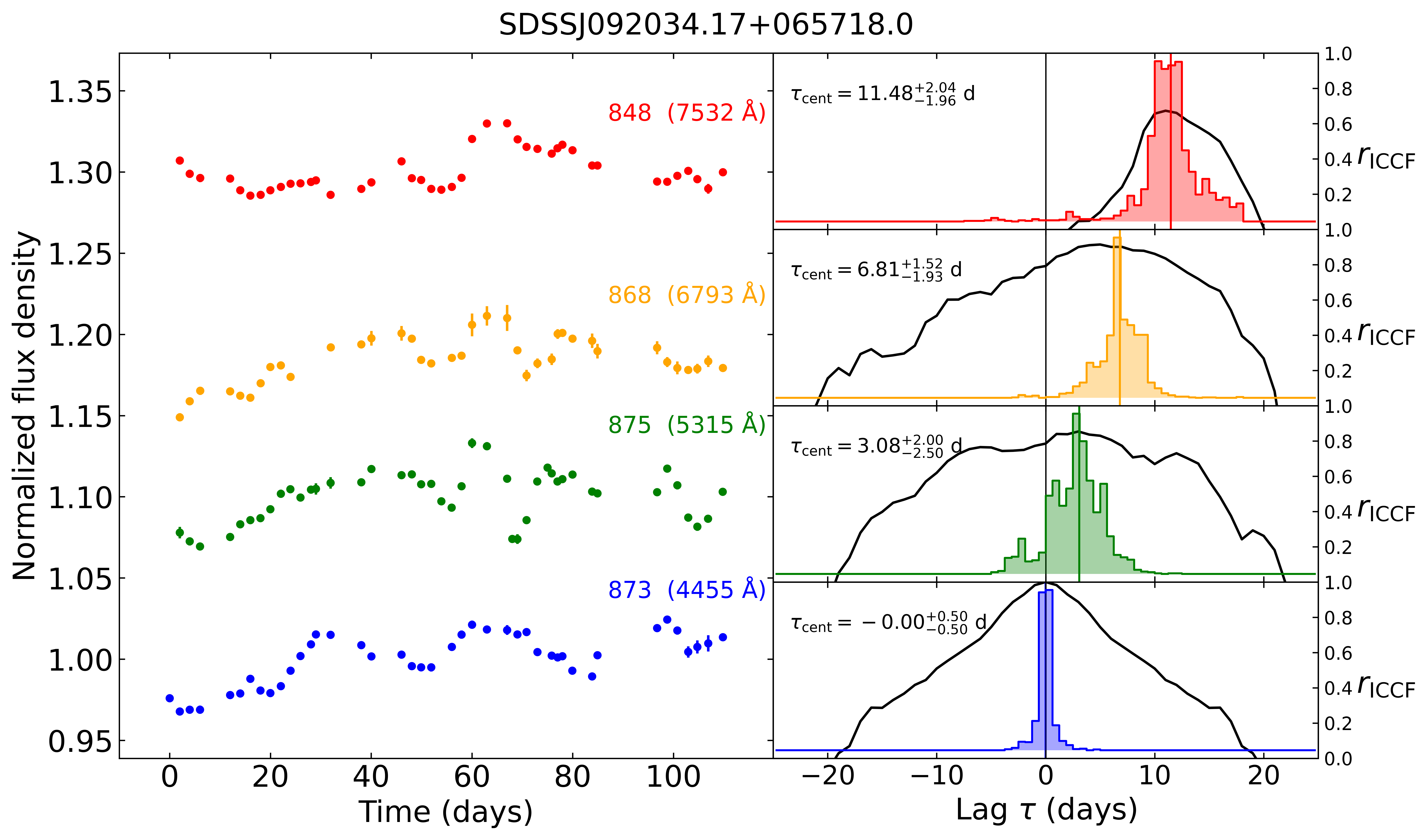}
\caption{Same as Fig.~\ref{fig:lcs_lags_SMSS}, but for J0920.
The WFI medium-band filters are 873, 875, 868, and 848, with the 873 band 
($\lambda_{\rm obs}=4455$\,\AA) used as the reference continuum band.}
\label{fig:lcs_lags_09}
\end{figure*}

\section{Results}
\label{res:lcs_discussion}

\subsection{Light curves}
\label{res:lcs}

The normalized light curves of J0529 and J0920 are shown in Figs.~\ref{fig:lcs_lags_SMSS} and \ref{fig:lcs_lags_09}, respectively.
To quantify their variability properties, we computed the observed RMS scatter, the mean photometric uncertainty, the ratio $Q \equiv \mathrm{RMS}/\langle\sigma\rangle$, and the noise-corrected fractional variability amplitude as introduced by \cite{1997ApJS..110....9R},
\begin{equation}
F_{\rm var} = \frac{\sqrt{S^2 - \langle \sigma^2 \rangle}}{\langle f \rangle},
\end{equation}
where $S^2$ is the sample variance of the measured fluxes in a given light curve, $\langle \sigma^2 \rangle$ is the mean squared photometric uncertainty, and $\langle f \rangle$ is the mean flux.
For the normalized light curves used here, $\langle f \rangle \simeq 1$. 
This quantity is widely used in RM analyses as a noise-corrected measure of the intrinsic variability amplitude, which is directly relevant to assess the reliability of lag measurements (e.g., \citealt{2024ApJ...962...67W,2026A&A...706A.176M}).
The ratio $Q$ provides a simple empirical diagnostic of whether the observed scatter exceeds that expected from measurement noise alone. 
Values of $Q\approx1$ indicate that the scatter is comparable to that expected from measurement noise, whereas $Q>1$ suggests excess variability beyond the noise level.
The resulting variability diagnostics are listed in Table~\ref{tab:variability_diagnostics}. 

Both quasars show significant intrinsic variability in all monitored bands, with $Q>1$.
However, their variability amplitudes differ markedly. 
In J0529 the variability is weak, with \(F_{\rm var}=0.28\)--0.51 per cent.
The reference band (867) has the largest variability significance, $Q=3.62$, reflecting both the largest RMS scatter and the smallest mean uncertainty (highest S/N) of the four bands.
Its noise-corrected fractional amplitude is nonetheless the smallest ($F_{\rm var}=0.28$ per cent): because the quasar is brightest in this band, the comparatively large absolute scatter corresponds to only a small fractional variability.
The remaining bands show $Q=1.56$--$1.92$, confirming that the observed fluctuations are still detected above the noise level.
These low amplitudes are consistent with the general trend that the most luminous quasars show relatively weak short-timescale continuum variability (e.g., \citealt{MacLeod2010}; \citealt{Simm2016}; \citealt{Stone2023}; \citealt{Tang2024}).

\begin{table*}[t]
\centering
\caption{Variability diagnostics for the continuum light curves (see Sect. \ref{res:lcs}).}
\label{tab:variability_diagnostics}
\begin{tabular}{lccccccc}
\hline\hline
Object & Filter & $\lambda_{\rm eff}$ (\AA) & $N$ & $\langle\sigma\rangle$ & RMS & $Q$ & $F_{\rm var}$ \\
\hline
SMSS\,J052915.80$-$435152.0 & 867 & 6463 & 67 & 0.00229 & 0.00830 & 3.62 & 0.00284 \\
 & 868 & 6793 & 66 & 0.00306 & 0.00589 & 1.92 & 0.00481 \\
 & 851 & 8159 & 65 & 0.00278 & 0.00434 & 1.56 & 0.00301 \\
 & 854 & 9148 & 65 & 0.00434 & 0.00686 & 1.58 & 0.00505 \\
\hline
SDSS\,J092034.17$+$065718.0 & 873 & 4455 & 42 & 0.00312 & 0.01572 & 5.04 & 0.01527 \\
 & 875 & 5315 & 44 & 0.00343 & 0.01304 & 3.80 & 0.01344 \\
 & 868 & 6793 & 41 & 0.00522 & 0.05555 & 10.64 & 0.05522 \\
 & 848 & 7532 & 42 & 0.00307 & 0.01303 & 4.24 & 0.01256 \\
\hline
\end{tabular}
\tablefoot{Columns give the object name, filter identifier, effective wavelength $\lambda_{\rm eff}$, number of epochs $N$, mean photometric uncertainty $\langle\sigma\rangle$, observed root-mean-square scatter (RMS), variability significance $Q \equiv {\rm RMS}/\langle\sigma\rangle$, and noise-corrected fractional variability amplitude $F_{\rm var}$.}
\end{table*}

By contrast, J0920 exhibits much stronger variability.
Its $Q$ values range from $3.80$ to $10.64$, indicating scatter well above the photometric uncertainties in all four bands. 
The largest variability is seen in the 868 band, where the RMS scatter reaches $5.6$ per cent and $Q=10.64$, while the other bands also show clear intrinsic variability.
The corresponding fractional variability amplitudes, \(F_{\rm var}=1.3\)--5.5 per cent, are substantially larger than in J0529 and indicate a strong continuum variability signal across the full dataset.

Despite this contrast in amplitude, both quasars show statistically significant continuum variability in all monitored bands, providing a solid basis for the inter-band lag analysis presented below.
Notably, the successful detection of sub-percent variability in J0529, one of the most luminous quasars known, demonstrates that CRM remains feasible even at the low-amplitude end expected for the most luminous quasars.

\subsection{Inter-band lag measurements}

We measured inter-band continuum lags using several standard RM estimators: the interpolated cross-correlation function (ICCF; \citealt{Gaskell1987}), the $Z$-transformed discrete correlation function (ZDCF; \citealt{1997ASSL..218..163A}), the von Neumann (VN) estimator (\citealt{2017ApJ...844..146C}), JAVELIN (\citealt{2011ApJ...735...80Z}), and the Gaussian Process Cross-Correlation (GPCC; \citealt{2023A&A...674A..83P}).
All methods were evaluated over a symmetric lag window of $\pm20$ days using identical lag grids. 
For the ICCF, ZDCF, and VN methods, uncertainties were estimated with the flux-randomization and random-subset selection (FR/RSS; \citealt{2004ApJ...613..682P}) procedure using $n_{\rm MC}=2000$ realizations. 
We adopt the centroid lag, $\tau_{\rm cent}$, as the primary estimator, computed over the region where the correlation coefficient exceeds $0.8 r_{\rm max}$. In the low-amplitude variability regime considered here, this approach provides a more stable measure of the delay than the peak lag.
For completeness, the peak lags are also listed in
Tables~\ref{tab:delays_all_methods_SMSS} and \ref{tab:delays_all_methods_SDSS}.
Peak and centroid lags agree within their uncertainties in all bands and in
both quasars, with rest-frame differences of at most $0.10$ days for J0529 and
$0.31$ days for J0920.
For J0529, lags are measured relative to the 867 band, while for J0920 the 873 band is used as the reference continuum band.
The resulting delays are listed in Tables~\ref{tab:delays_all_methods_SMSS} and \ref{tab:delays_all_methods_SDSS}. 
Given the overall consistency among the different methods, we show only the ICCF functions and their associated FR/RSS centroid distributions in Figs.~\ref{fig:lcs_lags_SMSS} and \ref{fig:lcs_lags_09}. 

In both quasars, the measured delays are positive relative to the reference band and increase monotonically with wavelength. 
The uncertainty ranges are broader for bands adjacent to the reference continuum and tighten as the wavelength separation increases, as expected from the larger spectral baseline available for the lag measurement. 
While J0529 shows a broader FR/RSS distribution in its longest-wavelength band, J0920 follows the general trend where the shortest wavelength separation yields the least constrained delay.
Overall, this lag hierarchy is consistent with thermal reprocessing in a stratified accretion disk, in which longer-wavelength emission originates from increasingly larger radii.
We discuss these results further in Sect. \ref{normalcompar}. 

\begin{table*}
\centering
\setlength{\tabcolsep}{15pt}
\renewcommand{\arraystretch}{1.2}
\caption{J0529 Rest-frame time-delay measurements obtained with different lag-recovery methods.}
\begin{tabular}{ccccccc}
\hline\hline
Filter & $\lambda_{\mathrm{obs}}$ [\AA] & $\tau_{\mathrm{ICCF}}$ & $\tau_{\mathrm{JAVELIN}}$ & $\tau_{\mathrm{ZDCF}}$ & $\tau_{\mathrm{VN}}$ & $\tau_{\mathrm{GPCC}}$ \\
\hline
867 & 6462 &
$0.00(0.00)_{-0.11}^{+0.11}$ &
$0.00_{-0.25}^{+0.21}$ &
$0.00(0.00)_{-0.10}^{+0.13}$ &
$0.00_{-0.20}^{+0.20}$ &
$0.00 \pm\,0.50$ \\

868 & 6793 &
$1.48(1.41)_{-0.18}^{+0.10}$ &
$1.54_{-0.66}^{+0.34}$ &
$1.38(1.42)_{-0.82}^{+0.77}$ &
$1.63_{-0.94}^{+0.87}$ &
$1.52 \pm\,0.68$ \\

851 & 8159 &
$1.72(1.81)_{-0.21}^{+0.30}$ &
$1.82_{-0.13}^{+0.12}$ &
$1.85(1.89)_{-0.23}^{+0.39}$ &
$1.83_{-0.43}^{+0.62}$ &
$1.79 \pm\,0.52$ \\

854 & 9148 &
$2.31(2.42)_{-0.60}^{+0.50}$ &
$2.27_{-0.25}^{+0.20}$ &
$2.35(2.39)_{-0.77}^{+0.52}$ &
$2.29_{-0.76}^{+0.68}$ &
$2.36 \pm\,0.57$ \\
\hline
\end{tabular}
\tablefoot{For the ICCF and ZDCF, the first value is the centroid lag and the
value in parentheses is the peak lag; the quoted uncertainties refer to the
centroid and were obtained with the FR/RSS method. GPCC delays are reported as the mode of the posterior distribution, with uncertainties corresponding to the 68\% highest posterior density (HPD) interval.}
\label{tab:delays_all_methods_SMSS}
\end{table*}

\begin{figure*}
\centering
\includegraphics[width=\columnwidth]{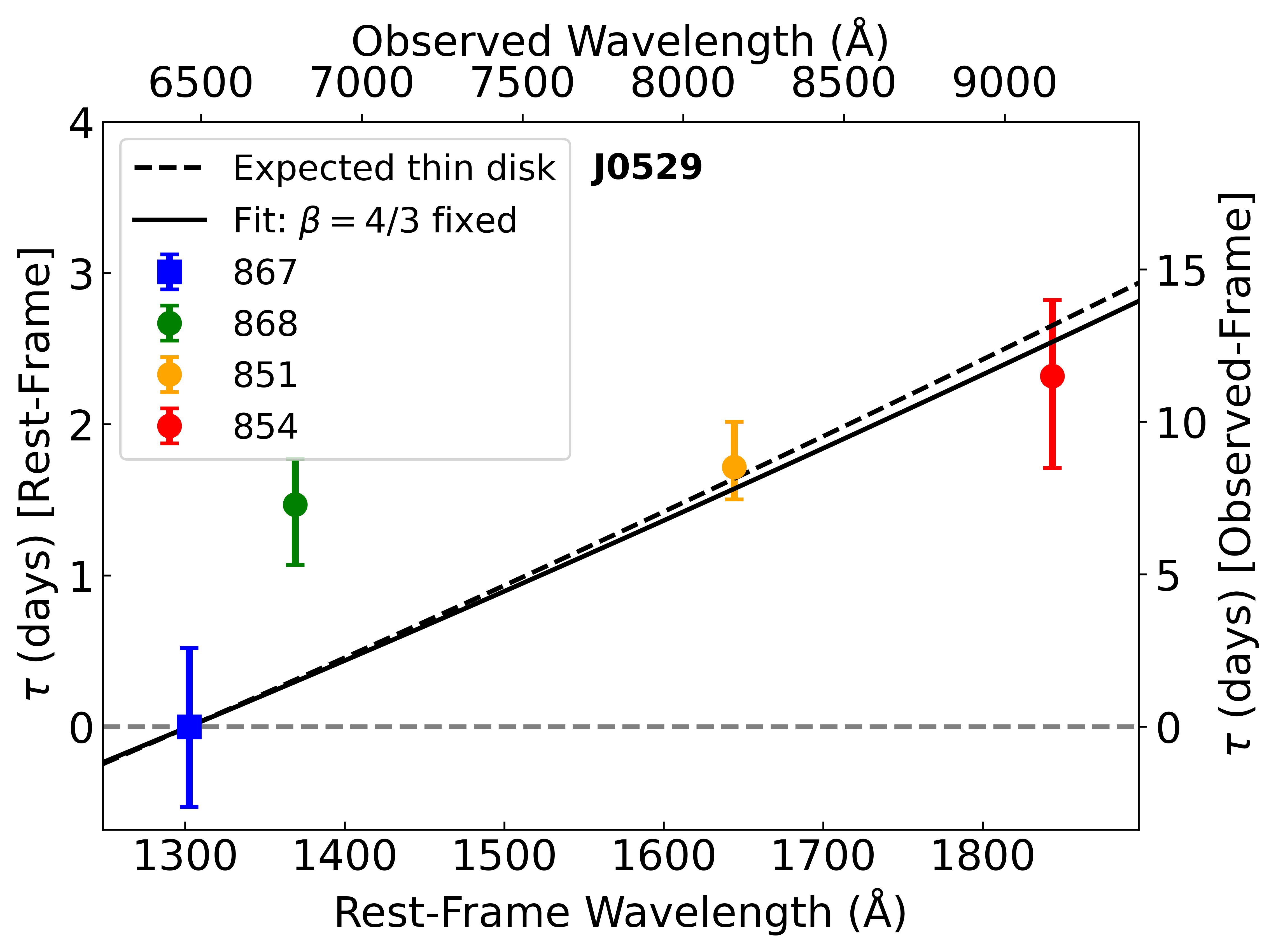}
\includegraphics[width=\columnwidth]{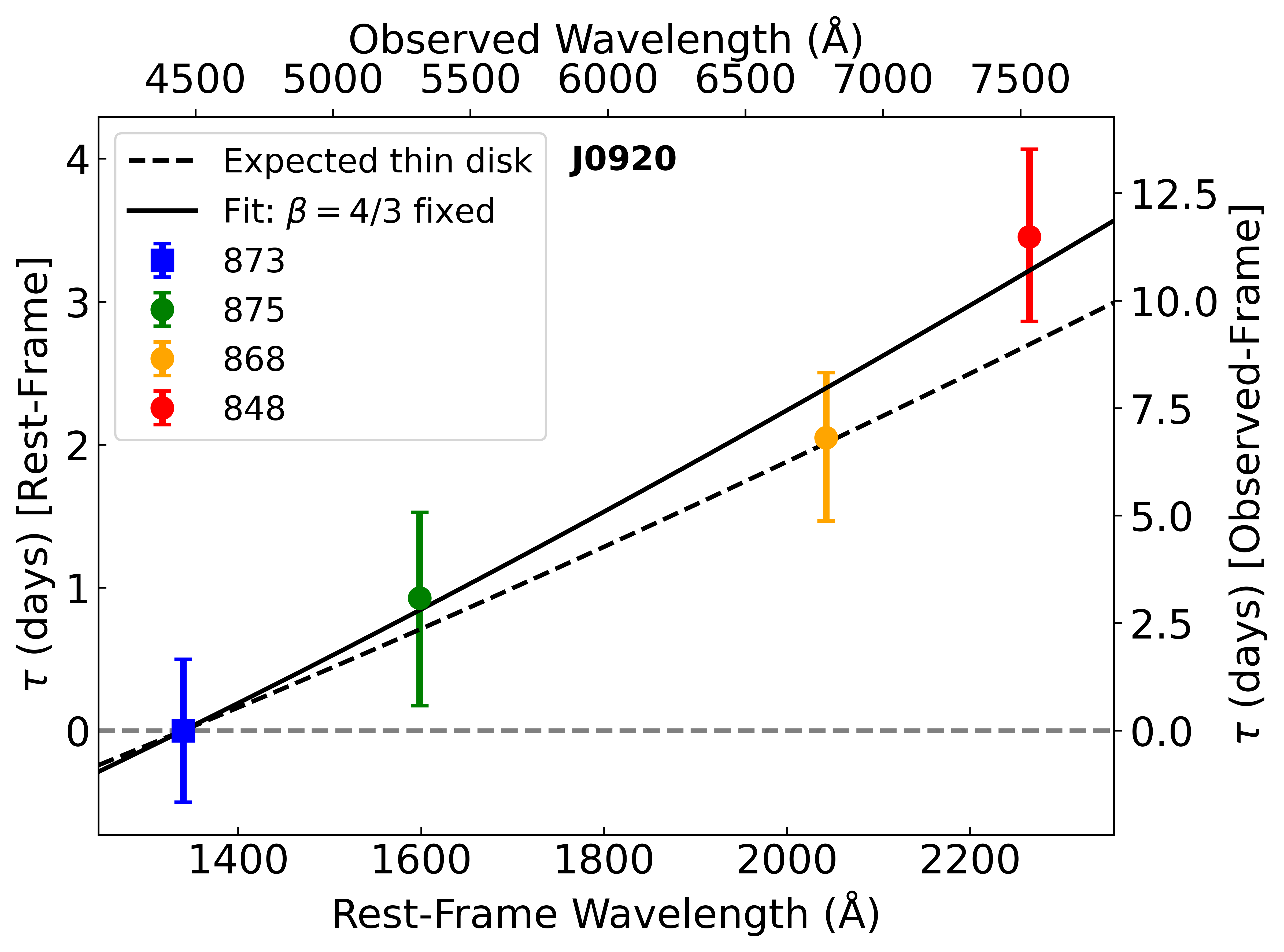}
\caption{Lag-wavelength relation for J0529 (left) and J0920 (right). The colored points show the centroid lags measured relative to the adopted reference continuum band using the ICCF method, with uncertainties estimated from the FR/RSS Monte Carlo distributions. The horizontal axis gives the rest-frame wavelength of each band, while the secondary axes indicate the corresponding observed-frame wavelength and lag. The solid black line shows the best-fit lag--wavelength relation with the slope fixed to the standard thin-disk value, $\beta=4/3$ (Eq.~\ref{eq:taulam_fixedbeta}), where the normalization $A$ represents the characteristic disk scale. The dashed line indicates the expected thin-disk prediction computed from the source luminosity assuming no disk inflation. In both objects, the measured lag spectrum is broadly consistent with the expected $\lambda^{4/3}$ trend, allowing a direct comparison between the observed disk size and the standard thin-disk expectation.}
\label{fig:delay_spectrum}
\end{figure*}

\subsection{Disk-scale normalization and comparison to thin-disk expectations}
\label{normalcompar}

The wavelength dependence of the continuum reverberation lag provides a direct probe of the radial temperature structure of the accretion disk (e.g., \citealt{1998ApJ...500..162C,Collier1998,Fausnaugh2016,Cackett2018}).
In the standard geometrically thin, optically thick disk model, the temperature profile $T\propto R^{-3/4}$ implies a lag--wavelength relation of the form

\begin{equation}
\tau_{\rm rest}(\lambda) =
A\left[
\left(\frac{\lambda}{\lambda_0}\right)^{\beta}-1
\right],
\label{eq:taulam_fixedbeta}
\end{equation}
where $\lambda_0$ is the reference continuum wavelength and $\beta=4/3$ for a standard thin disk.
This parameterization is commonly used in continuum reverberation studies to describe the observed lag spectrum (e.g., \citealt{2023ApJ...958..195C,2024ApJ...973..152E,2026A&A...706A.176M}).

To place the measured normalization in a physical context, we recall the corresponding thin-disk expectation.
The expected continuum-emitting radius as a function of wavelength follows from the standard temperature profile,
\begin{equation}
T(R)=\left(\frac{3GM_{\rm BH}\dot{M}}{8\pi\sigma R^3}\right)^{1/4},
\label{eq:Temp}
\end{equation}
where $M_{\rm BH}$ is the black-hole mass, $\dot{M}$ the mass accretion rate, and $\sigma$ the Stefan--Boltzmann constant.
The radius associated with a given wavelength is obtained by equating the disk temperature to the photon energy scale
\begin{equation}
k_{\rm B}T(R_\lambda)=\frac{hc}{X\,\lambda},
\label{eq:Rlambda}
\end{equation}
where $X$ accounts for the response-weighted emission radius of the disk.
Combining these expressions yields the scaling relation
\begin{equation}
R_\lambda \propto
\left(\frac{3GM_{\rm BH}\dot{M}}{8\pi\sigma}\right)^{1/3}
\left(\frac{Xk_{\rm B}\lambda}{hc}\right)^{4/3},
\label{eq:frac}
\end{equation}
which provides the corresponding radial scaling of the standard disk,
$R_\lambda\propto(M_{\rm BH}\dot{M})^{1/3}\lambda^{4/3}$.

Because our filter set provides only limited wavelength leverage, we fix the slope to the thin-disk value $\beta=4/3$ and fit only for the normalization $A$, which sets the characteristic disk scale, $R_0=cA$.
The resulting rest-frame lag--wavelength relation is shown in Figure~\ref{fig:delay_spectrum}.
In both quasars, the measured centroid lags increase with wavelength and are well described by the fixed-slope thin-disk relation.

The uncertainty in $A$ was estimated by Monte Carlo sampling of the FR/RSS centroid-lag distributions.
In each realization, we draw one centroid lag $\tau_{{\rm cent},i}$ for each filter $i$ relative to the reference band, convert the delays to the rest frame, and solve for the corresponding normalization $A$.
Repeating this procedure yields a posterior distribution for $A$ that naturally captures the asymmetric and non-Gaussian lag uncertainties.

For J0529, we find $A = R_{\rm AD} = 4.35^{+0.78}_{-0.91}\ {\rm days}$, where $R_{\rm AD}$ is the characteristic disk scale.
We compare the measured normalization with the corresponding value predicted by standard thin-disk theory for the fiducial single-epoch mass of J0529.
Using the same functional form for the lag--wavelength relation we obtain an expected normalization $A_{\rm pred}=4.50$\,days.
The ratio between the measured and predicted disk scales is
\begin{equation}
f = \frac{A}{A_{\rm pred}} = 0.97^{+0.17}_{-0.20}.
\end{equation}
Thus, the measured disk scale is consistent with the nominal thin-disk expectation within the uncertainties\footnote{The uncertainty in $f$ reflects only the FR/RSS uncertainty in the measured disk-scale normalization $A$, while the theoretical prediction $A_{\rm pred}$ is treated as fixed.
Systematic uncertainties associated with the black-hole mass, bolometric luminosity, radiative efficiency, and disk--wavelength conversion factor are not included.}.
Adopting the GRAVITY+ dynamical mass instead would lower $A_{\rm pred}$ and raise $f$ to $\approx2.8$ (see Sect.~\ref{res:discussion}).

\begin{table*}
\centering
\setlength{\tabcolsep}{15pt}
\renewcommand{\arraystretch}{1.2}
\caption{J0920 Rest-frame time-delay measurements obtained with different lag-recovery methods.}
\begin{tabular}{ccccccc}
\hline\hline
Filter & $\lambda_{\mathrm{obs}}$ [\AA] & $\tau_{\mathrm{ICCF}}$ & $\tau_{\mathrm{JAVELIN}}$ & $\tau_{\mathrm{ZDCF}}$ & $\tau_{\mathrm{VN}}$ & $\tau_{\mathrm{GPCC}}$ \\
\hline
873 & 4455 &
$0.00(0.00)_{-0.15}^{+0.15}$ &
$0.00_{-0.22}^{+0.43}$ &
$0.00(0.00)_{-0.17}^{+0.18}$ &
$0.00_{-0.30}^{+0.32}$ &
$0.00 \pm\,0.40$ \\

875 & 5315 &
$0.93(0.91)_{-0.75}^{+0.60}$ &
$0.95_{-0.25}^{+0.23}$ &
$1.04(0.97)_{-0.66}^{+0.51}$ &
$1.02_{-0.60}^{+0.40}$ &
$0.91 \pm\,0.46$ \\

868 & 6793 &
$2.04(1.73)_{-0.58}^{+0.45}$ &
$2.15_{-0.48}^{+0.42}$ &
$2.11(2.00)_{-0.75}^{+0.64}$ &
$2.11_{-1.23}^{+1.20}$ &
$1.95 \pm\,0.56$ \\

848 & 7532 &
$3.45(3.23)_{-0.78}^{+0.81}$ &
$3.34_{-0.56}^{+0.75}$ &
$3.53(3.24)_{-0.81}^{+0.94}$ &
$3.61_{-1.10}^{+1.21}$ &
$3.56 \pm\,0.73$ \\
\hline
\end{tabular}
\tablefoot{For the ICCF and ZDCF, the first value is the centroid lag and the
value in parentheses is the peak lag; the quoted uncertainties refer to the
centroid and were obtained with the FR/RSS method. GPCC delays are reported as the mode of the posterior distribution, with uncertainties corresponding to the 68\% HPD interval.}
\label{tab:delays_all_methods_SDSS}
\end{table*}

For J0920 we find $A = R_{\rm AD} = 3.15^{+0.50}_{-0.48}\ {\rm days}$.
The predicted thin-disk normalization for the fiducial GRAVITY dynamical mass is $A_{\rm pred}=2.66$\,days, yielding
\begin{equation}
f = \frac{A}{A_{\rm pred}} = 1.18^{+0.19}_{-0.18}.
\end{equation}
In this case, the measured disk scale is modestly larger than the nominal thin-disk prediction, but remains consistent with it within the $1\sigma$ uncertainties.
Conversely, adopting the larger single-epoch masses would raise $A_{\rm pred}$ and lower $f$ below unity ($f\approx0.5$--$0.7$; Sect.~\ref{res:discussion}).

Figure~\ref{fig:delay_spectrum} compares the measured lag spectra with the nominal thin-disk prediction for $f=1$.
The two quasars lie close to the standard thin-disk expectation: J0529 is consistent with unity under its single-epoch mass, while J0920 shows at most a mild excess under its dynamical mass.
Neither source requires the factor of $\sim2$--$3$ disk inflation often reported in lower-redshift continuum reverberation studies.
This result is noteworthy, even though it is currently based on only two objects.
One possible interpretation is that ground-based CRM of high-redshift quasars samples relatively short rest-frame wavelengths, where contamination from diffuse BLR continuum emission is expected to be weaker (\citealt{Pozo2025}).
Photoionization calculations indicate that the BLR continuum contributes more strongly at longer wavelengths, particularly above $\sim2000$\,\AA\ (see, e.g., Fig. 2b of \citealt{korista2001}; Fig. 2 of \citealt{2022MNRAS.509.2637N}; Fig. 4 of \citealt{jaiswal2025}; or Fig. 7 in \citealt{2026A&A...706A.176M}).
If this interpretation is correct, continuum monitoring of high-redshift quasars may prove easier to interpret than in lower-redshift systems, and future wide-field surveys could provide especially valuable datasets for accretion-disk studies.

A related consideration is that this monitoring also probes much shorter rest-frame timescales than typical low-redshift campaigns.
The large black-hole masses of our targets mean that the measured rest-frame lags of $\sim3$ days would correspond to $\lesssim0.3$ days in a $10^{7}\,M_\odot$ system, and our $\sim150$-day observed-frame baseline would likewise correspond to a far shorter campaign in a nearby, lower-mass AGN.
More importantly, the interferometric BLR sizes of J0529 and J0920 ($\sim1190$ and $\sim360$ light-days, respectively) greatly exceed our monitoring baseline, so the campaign cannot sample the timescales on which diffuse BLR continuum emission would respond.
Our measurement is therefore analogous to a frequency-resolved lag analysis restricted to high frequencies, where the contribution of an extended reprocessor is suppressed.
Frequency-resolved studies of nearby AGN find that high-frequency lags are generally consistent with standard thin-disk sizes, whereas the excess normalization emerges at low frequencies (\citealt{Cackett2022,Lewin2023,Drewes2026}).
The agreement with thin-disk expectations found here may therefore reflect, at least in part, the restricted range of timescales probed rather than an intrinsic difference between high- and low-redshift accretion disks.

\begin{figure}
\centering
\includegraphics[width=\columnwidth]{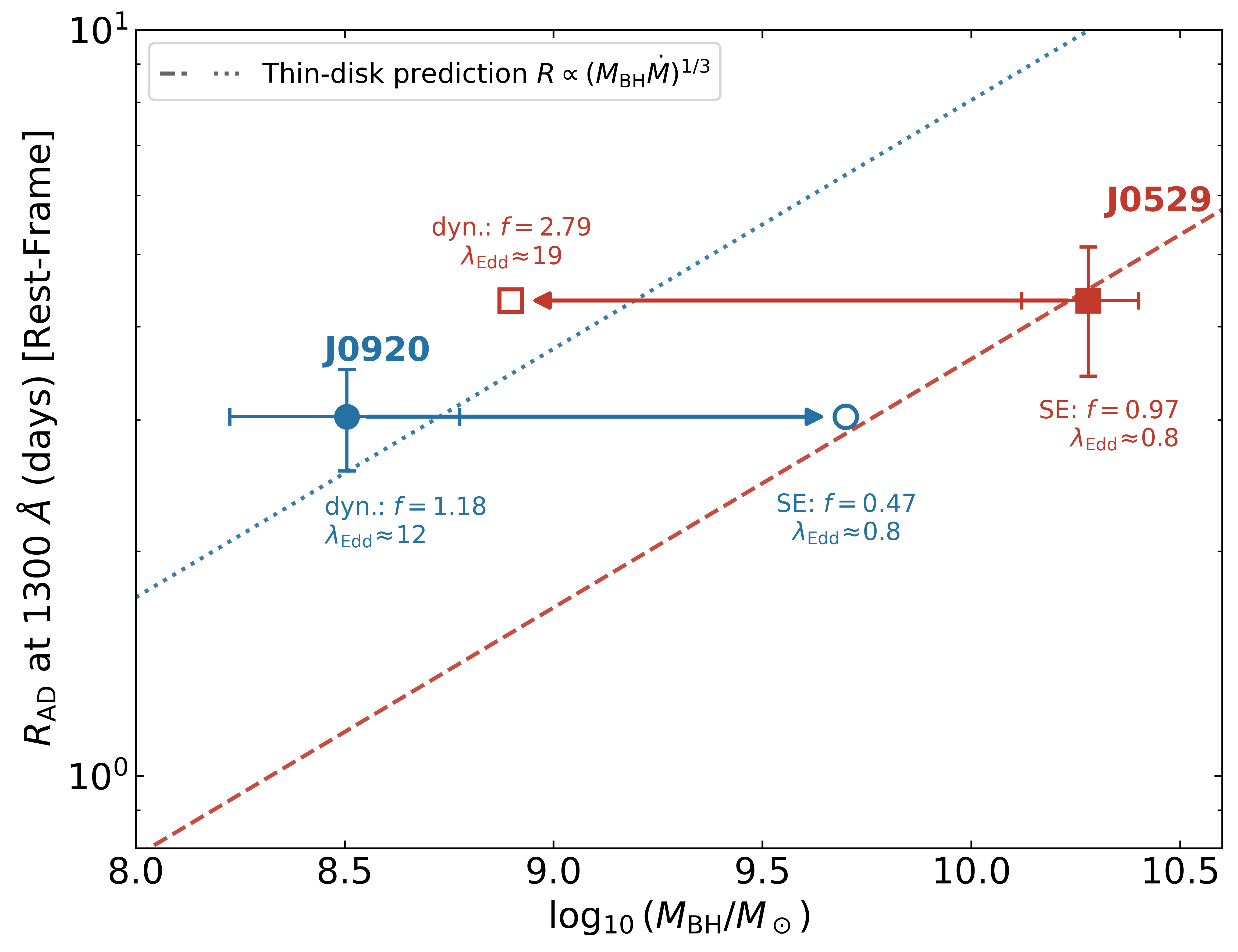}
\caption{Accretion-disk size at 1300\,\AA\ as a function of black-hole mass for J0529 and J0920. 
Filled symbols mark each quasar at its adopted mass (single-epoch for J0529, GRAVITY
dynamical for J0920); open symbols and horizontal arrows show the same measured
size at the alternative mass estimate (the dynamical mass for J0529, the
single-epoch mass for J0920). Dashed red (J0529) and dotted blue (J0920) lines are each quasar's thin-disk
prediction, $R\propto(M_{\rm BH}\dot{M})^{1/3}$, evaluated at its fixed $L_{\rm
bol}$. Each label gives the mass type (SE or dyn.), the disk-size ratio $f=R_{\rm AD}/A_{\rm pred}$ and the
Eddington ratio at each mass. Horizontal 
bars show the uncertainty on the adopted mass and vertical bars the 16th--84th 
percentile range of $R_{\rm AD}$; all error bars are statistical only and do 
not include systematic uncertainties.}
\label{fig:disk_mass}
\end{figure}

Figure~\ref{fig:disk_mass} shows the inferred disk sizes of both quasars,
rescaled to a common rest-frame wavelength of 1300\,\AA, in the
black-hole-mass versus disk-size plane. Each quasar is plotted at its adopted
black-hole mass (filled symbols), the single-epoch mass for J0529 and the
GRAVITY dynamical mass for J0920, while the open symbols and horizontal
arrows show the same measured size re-plotted at the alternative mass
estimate. Because the disk size is an observable, adopting a different
black-hole mass produces only a horizontal shift in this plane. The dashed
line associated with each quasar is its thin-disk prediction, adopting $\eta=0.1$ and a conversion factor of $X=2.49$\footnote{The value of $X=2.49$ corresponds to a flux-weighted mean radius $\langle R \rangle = {\int_{R_0}^{\infty} B(T(R))R^2\,dR}/{\int_{R_0}^{\infty} B(T(R))R\,dR}$, where $B(T(R))$ is the Planck function and we assume a temperature profile $T\propto R^{-3/4}$ (\citealt{Fausnaugh2016,Edelson2017,Pozo2019}).}.
At their adopted masses both quasars lie close to the nominal
thin-disk prediction ($f=0.97$ for J0529 and $f=1.18$ for J0920). Adopting the
alternative mass instead would place J0529 well above its prediction (an
inflated disk, $f\approx2.8$, in the super-Eddington regime implied by the
smaller dynamical mass) and J0920 below its prediction (an under-sized disk,
$f\approx0.5$). The measured agreement with thin-disk theory is thus robust at
the adopted masses but, as expected, depends on the assumed black-hole mass.

These measurements establish the characteristic scale of the UV
continuum-emitting region in both quasars. In the following section, we
compare these disk sizes with the independently constrained BLR sizes from
GRAVITY to situate the accretion disk and line-emitting region within a
unified structural framework.

\subsection{Accretion-disk and broad-line region scales}
\label{blrsizecompar}

Recent interferometric observations with GRAVITY provide independent constraints on the BLR in both quasars. 
For J0529, GRAVITY+ resolves the H$\beta$ BLR and finds that the line-emitting gas is dominated by a strong outflow component, which contributes $\sim83\%$ of the total line flux (\citealt{Gravity2026}). 
Modeling yields a characteristic H$\beta$ radius of about $1.0$\,pc ($\sim1190$ light-days) and a dynamical black-hole mass of $M_{\rm BH}\simeq8\times10^{8}\,M_\odot$. 
For J0920, GRAVITY resolves the H$\alpha$ BLR and infers a mean radius of roughly $0.3$\,pc ($\sim360$ light-days). 
In this case, the line-emitting gas is consistent with a moderately inclined rotating disk, with a corresponding dynamical black-hole mass of $M_{\rm BH}\simeq3\times10^{8}\,M_\odot$ (\citealt{Abuter2024});
we adopt this as our fiducial mass and discuss the alternative single-epoch estimates in Sect.~\ref{res:discussion}.

\begin{figure}
\centering
\includegraphics[width=\columnwidth]{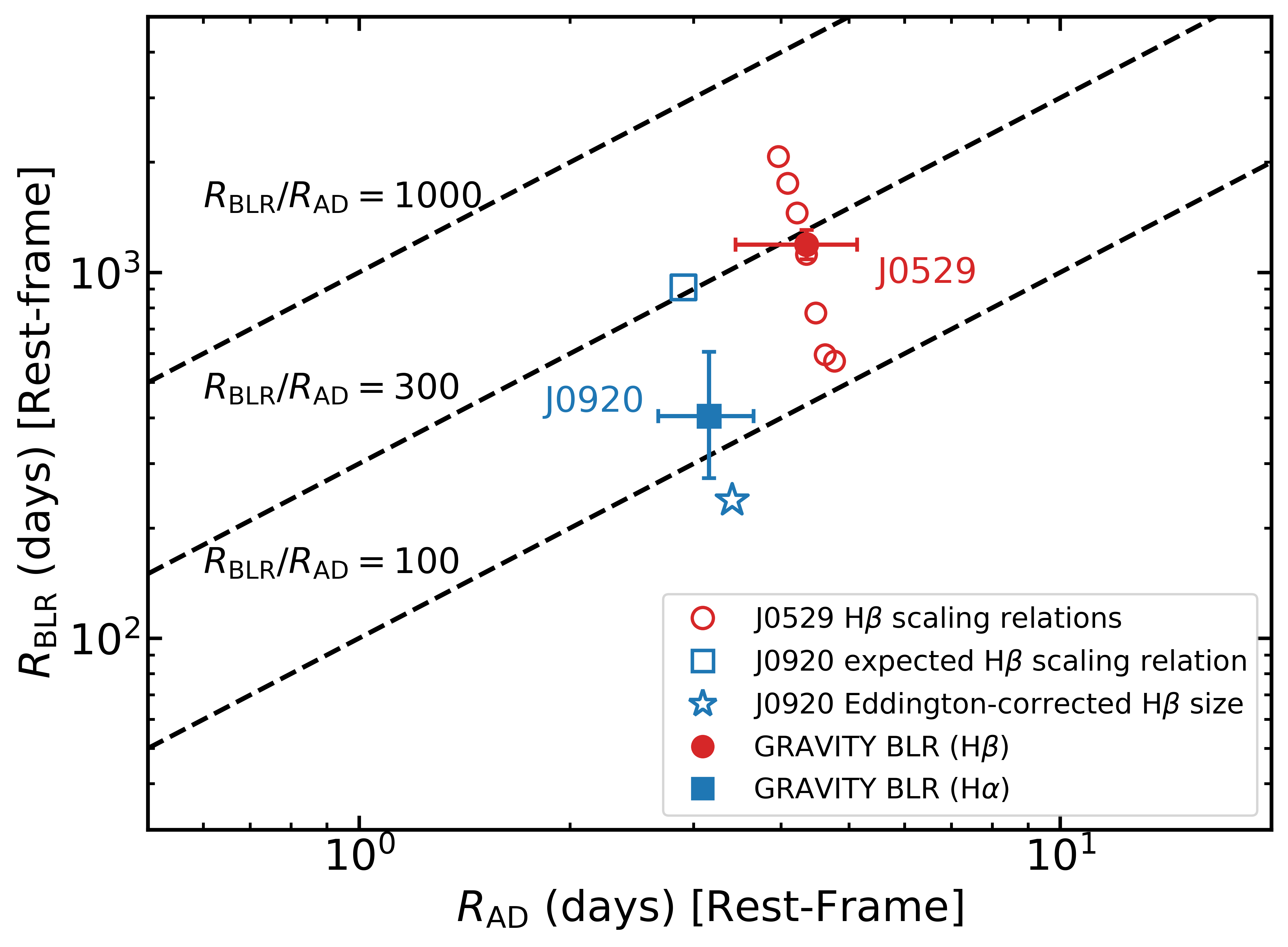}
\caption{Comparison between the characteristic accretion-disk scale and the BLR size for J0529 and J0920. The filled symbols denote the BLR sizes measured with GRAVITY. 
For J0529, the open red circles indicate BLR sizes from literature H$\beta$ $R$--$L$ relations (see Table 2 in \citealt{Gravity2026} and references therein). 
For J0920, the open blue square marks the H$\beta$ size expected from the local $R$--$L$ relation, while the open blue star shows the Eddington-ratio-corrected H$\beta$ estimate, both taken from \citet{Abuter2024}. 
Dashed diagonal lines indicate constant ratios $R_{\rm BLR}/R_{\rm AD}$.
}
\label{fig:blr_vs_disk}
\end{figure}

Figure~\ref{fig:blr_vs_disk} compares the UV accretion-disk sizes derived in this work with the BLR sizes measured via GRAVITY.
These interferometric BLR sizes are measured in different lines---H$\beta$ for J0529 (GRAVITY+) and H$\alpha$ for J0920 (GRAVITY).
For J0529, the UV disk scale of $R_{\rm AD}=4.35^{+0.78}_{-0.91}$ light-days, implies a radius ratio of

\begin{equation}
\frac{R_{\rm BLR}}{R_{\rm AD}}\approx270.
\end{equation}

For J0920, the disk size $R_{\rm AD}=3.15^{+0.50}_{-0.48}$ light-days yields

\begin{equation}
\frac{R_{\rm BLR}}{R_{\rm AD}}\approx115.
\end{equation}

Consequently, in both quasars the Balmer-line BLR is more than two orders of magnitude larger than the UV continuum-emitting accretion disk.
Because  H$\alpha$ typically traces a somewhat larger radius than H$\beta$, these two  ratios are not strictly on a common footing; placing both objects on the same  line would, if anything, widen the contrast between them rather than reduce it.
For comparison, Fig.~\ref{fig:blr_vs_disk} also includes BLR size estimates derived from various Balmer-line radius-luminosity ($R-L$) relations reported in the literature. 
In J0529, these relations predict BLR radii ranging from $\sim0.5$ to $\sim1.7$ pc ($\sim600$–$2000$ light-days); the interferometric H$\beta$ measurement lies within this spread.
In J0920, by contrast, the GRAVITY H$\alpha$ radius is smaller than the size expected from the local H$\beta$-based $R-L$ relation by approximately a factor of two. 
However, the Eddington-ratio-corrected H$\beta$ estimate of \citet{Abuter2024}, which uses the $R_{\rm Fe}$-dependent $R$--$L$ relation of \citet{2019ApJ...886...42D}, yields $\sim0.2$ pc, much more consistent with the interferometric result.
This finding supports the hypothesis that high-Eddington accretion can lead to a more compact BLR than predicted by standard $R-L$ scaling, potentially due to effects of an altered ionizing SED or radiation pressure (\citealt{2019ApJ...886...42D,2021A&A...650A.154P}).

It is important to note that this comparison does not represent a direct line–continuum reverberation pair.
Our continuum measurements probe the UV disk at $\lambda_{\rm rest}\simeq1300$\,\AA\, whereas the interferometric constraints refer to low-ionization Balmer-line emitting regions.
This comparison is therefore structural, linking the characteristic size of UV-emitting disk to the outer BLR.
In this framework, the larger ratio found for J0529 primarily reflects the extended H$\beta$ geometry inferred from GRAVITY+, while the smaller ratio in J0920 corresponds to its more compact interferometric BLR.

\begin{figure}
\centering
\includegraphics[width=\columnwidth]{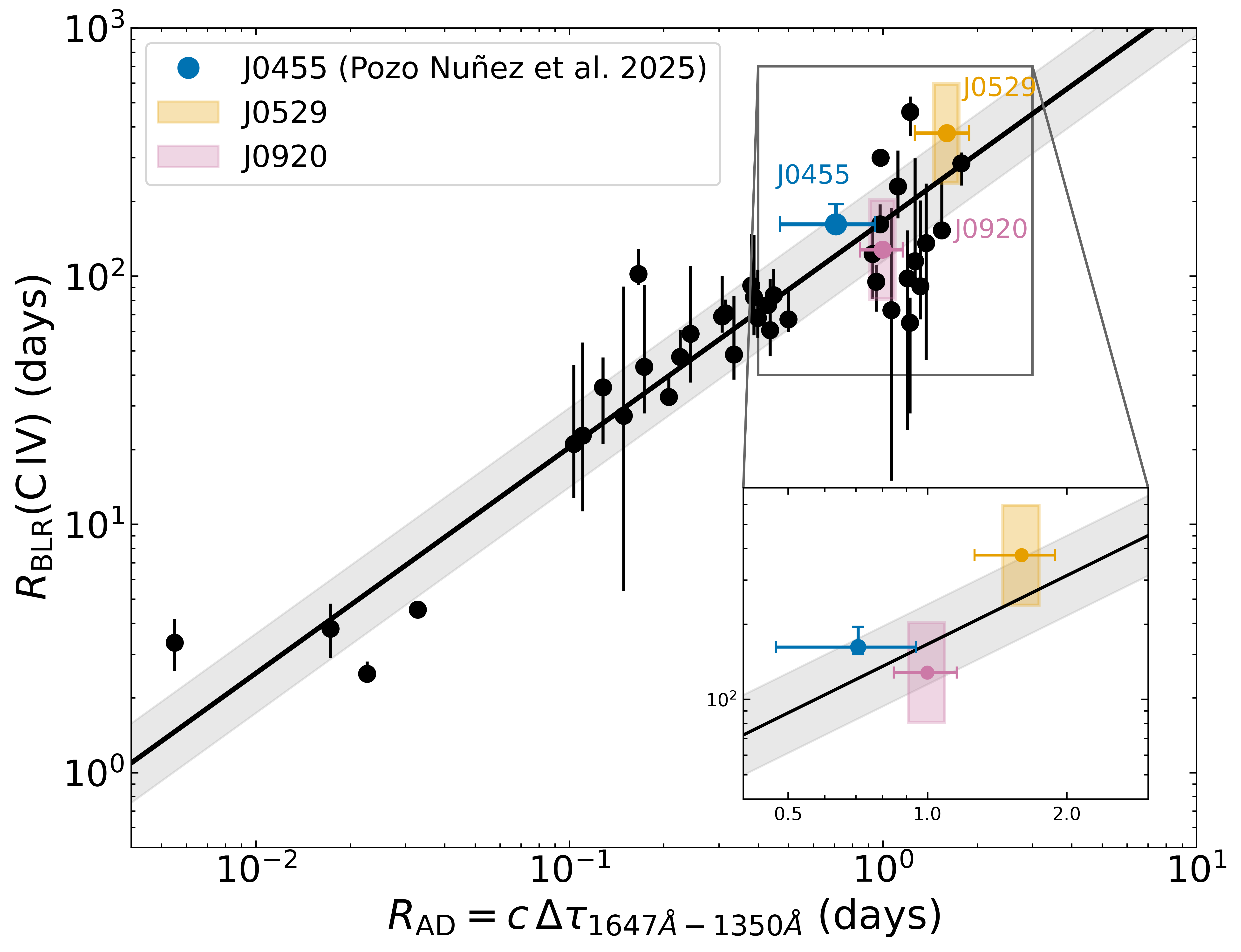}
\caption{C\,IV BLR size as a function of the continuum size relevant for the CIV emitting region, defined here as $R_{\rm AD}=c\,\Delta\tau_{1647-1350}$. 
The solid black line and gray shaded band show the size--size relation proposed by \citet{Panda2024} and its intrinsic scatter. 
The blue point marks QSO\,J0455$-$4216, for which both the continuum delay and the C\,IV BLR size are available from reverberation mapping. 
The orange and green shaded regions denote the inferred CIV BLR intervals for J0529 and J0920, respectively, obtained by combining the measured continuum sizes with C\,IV BLR ranges from the GRAVITY Balmer-line BLR sizes under the assumption
that the high-ionization C\,IV emitting region lies interior to the Balmer-line region (see text). No C\,IV lags are measured for these two quasars: the shaded regions are inferred from scaling relations and should be regarded as order-of-magnitude estimates rather than measurements. The inset shows an enlarged view of the region occupied by the three high-redshift quasars for clarity.}
\label{fig:civ_blr_ad_three}
\end{figure}

As an exploratory extension, we compare our results with the C\,IV BLR–disk relation proposed by \citet{Panda2024}. In this model, the relevant continuum scale is the differential size $R_{\rm AD}=c\,\Delta\tau_{1647-1350}$, rather than the absolute disk scale at a single wavelength.
Within this framework, the C\,IV BLR is typically $\sim150$ times larger than the corresponding UV continuum scale.
Figure~\ref{fig:civ_blr_ad_three} places our quasars in this context alongside QSO J0455-4216, for which both the continuum delay and the C\,IV BLR size are available from reverberation mapping \citep{Pozo2025}.
Since direct C\,IV BLR measurements are currently unavailable for J0529 and J0920, we instead show inferred ranges obtained from scaling relations.
Their horizontal positions are derived from our measured lag-spectrum normalizations, while the vertical ranges are inferred from the GRAVITY Balmer-line sizes, assuming the high-ionization C\,IV region lies interior to the Balmer-line region.

For J0920, we first convert the measured H$\alpha$ size to an approximate H$\beta$ scale adopting the empirical ratio $\tau({\rm H}\alpha)/\tau({\rm H}\beta)\simeq1.54$ (\citealt{Bentz2010}). 
We then assume an illustrative range of $R_{\rm BLR}({\rm C\,IV})=(0.5$--$1.0)\,R_{\rm BLR}({\rm H}\beta)$, motivated by multi-line reverberation studies indicating that C\,IV lags are typically shorter than those of H$\beta$. 
The exact ratio remains uncertain, with recent OzDES results favoring $\sim0.8$ (\citealt{McDougall2025}) compared to earlier estimates of $\sim0.5$; the resulting ranges therefore span roughly a factor of two in $R_{\rm BLR}({\rm C\,IV})$.

These comparisons support a stratified structural model where the UV-emitting disk occupies the innermost few light-days, while the Balmer-line BLR lies at radii two to three orders of magnitude larger.
The inferred C\,IV regions of both quasars fall close to the $R_{\rm BLR}({\rm C\,IV})$--$R_{\rm AD}$ relation of \citet{Panda2024} (Fig.~\ref{fig:civ_blr_ad_three}). We emphasize, however, that these are order-of-magnitude estimates rather than measurements: no C\,IV lag is measured here, and the inferred ranges rest on a chain of scaling relations linking the GRAVITY Balmer-line sizes to an assumed C\,IV radius. Given the width of these ranges and the intrinsic scatter of the relation, this agreement should be regarded as a consistency check rather than as independent support for the relation.
If confirmed with directly measured C\,IV lags, such a relation could serve as a promising black hole mass estimator using CRM when combined with a broad-line velocity measurement from a single-epoch spectrum, provided that it is robustly calibrated over a sufficiently wide luminosity range. \citet{Mandal25} demonstrated such potential using the $R_{\rm BLR}({\rm H\beta})$--$R_{\rm AD}$ relation, primarily for local AGNs. In this context, the analogous $R_{\rm BLR}({\rm C\,IV})$--$R_{\rm AD}$ relation is expected to be particularly valuable for high-redshift quasars, for which traditional spectroscopic RM campaigns are observationally expensive and time-intensive.
A more detailed physical interpretation, based on dedicated light-curve and SED modeling, will be presented in a forthcoming publication.

\section{Discussion}
\label{res:discussion}

\subsection{Disk-size normalization in the context of GRAVITY BLR constraints}
\label{res:disksize}
The lag spectra of both quasars are consistent with a thin-disk relation of fixed slope $\beta=4/3$, but the physical interpretation of the normalization depends heavily on the adopted black-hole mass and accretion rate.
From Eq.~\ref{eq:frac}, the predicted disk scale at a fixed wavelength scales as  $A_{\rm pred}(\lambda)\propto\left(M_{\rm BH}\dot{M}\right)^{1/3}$, or equivalently

\begin{equation}
A_{\rm pred}(\lambda)\ \propto\ \left(\frac{M_{\rm BH}\,L_{\rm bol}}{\eta}\right)^{1/3}.
\label{eq:Apred_scaling_Lbol}
\end{equation}
assuming the accretion rate is expressed via the bolometric luminosity.
Although this dependence is only to the one-third power, an order-of-magnitude uncertainty in $M_{\rm BH}$ or $L_{\rm bol}$ shifts the predicted disk scale by a factor of $\sim2.15$.
This shift is sufficient to determine whether a source appears consistent with standard thin-disk theory or instead requires an "inflated" disk model.

This is particularly relevant for J0529. 
The GRAVITY+ analysis indicates that its Balmer-line BLR is dominated by a strong outflow, yielding a dynamical black-hole mass substantially smaller than typical single-epoch virial estimates.
In such cases, standard single-epoch calibration may be biased, as the observed line widths might not trace a predominantly virialized region. 
This smaller mass estimates implies a significantly higher Eddington ratio ($\lambda_{\rm Edd}\sim6$--$19$), suggesting a super-Eddington regime characterized by geometrically thick inner flows and radiation-driven winds.
Under the single-epoch mass assumption adopted in Section \ref{blrsizecompar}, our measured UV disk size aligns with the nominal thin-disk prediction.
However, if the smaller GRAVITY+ dynamical mass is adopted, the predicted thin-disk size decreases substantially, and the observed disk would appear "inflated" by a factor of $f\approx2.8$--consistent with the disk-size excess frequently reported in lower-redshift AGN (e.g., \citealt{Edelson2019,HernandezSantisteban2020,Buitrago25}). 
While this comparison does not uniquely favor one mass estimate over the other, it demonstrates that the UV continuum-emitting region provides an independent physical scale to assess BLR-based mass and accretion-rate estimates.

An analogous but inverted situation applies to J0920. \citet{Abuter2024} 
report several black-hole mass estimates for this quasar spanning more than 
an order of magnitude. Single-epoch virial masses based on the C\,IV line 
give $\log(M_{\rm BH}/M_\odot)\approx9.7$; however, the C\,IV profile is 
strongly blueshifted and asymmetric, indicating a substantial non-virial 
(outflow) contribution that biases this estimate upward, and a 
blueshift-corrected C\,IV mass is $\log(M_{\rm BH}/M_\odot)\approx8.7$. 
The single-epoch H$\beta$ and H$\alpha$ masses are 
$\log(M_{\rm BH}/M_\odot)=9.24\pm0.47$ and $8.94\pm0.48$, respectively, the 
former lying $\sim0.7$\,dex above the dynamical value largely because the 
standard $R$--$L$ relation overestimates the BLR radius in this high-Eddington 
source. The GRAVITY spectro-interferometric (dynamical) mass, 
$\log(M_{\rm BH}/M_\odot)=8.51^{+0.27}_{-0.28}$ 
($M_{\rm BH}\simeq3.2\times10^{8}\,M_\odot$), provides the most direct 
measurement \citep{Abuter2024}: it is obtained self-consistently from the spatially resolved 
H$\alpha$ kinematics without recourse to a virial scaling factor, and it 
carries the smallest uncertainty. When the single-epoch H$\beta$ radius is 
corrected for the Eddington ratio, the resulting mass 
($\log(M_{\rm BH}/M_\odot)\approx8.6$) agrees with the dynamical value to 
within $0.1$\,dex.

Crucially, our continuum reverberation measurement provides
independent support for the dynamical mass. Because 
$A_{\rm pred}\propto(M_{\rm BH}L_{\rm bol})^{1/3}$ 
(Eq.~\ref{eq:Apred_scaling_Lbol}), adopting the larger single-epoch masses 
would raise the predicted thin-disk scale and drive the observed-to-predicted 
ratio well below unity: the uncorrected C\,IV mass would yield $f\approx0.5$ 
and the single-epoch H$\beta$ mass $f\approx0.7$, implying a UV 
continuum-emitting region a factor of $\sim2$ smaller than the 
thin-disk prediction. Such an under-sized disk would be difficult to 
reconcile with reprocessing in a standard disk and runs opposite to the 
disk-size excess commonly reported at lower redshift. In contrast, the 
GRAVITY dynamical mass, and the Eddington-corrected H$\beta$ mass, which 
is consistent with it, yields $f=1.18^{+0.19}_{-0.18}$, in agreement with 
nominal thin-disk expectations.

Together, these cases show that disk size, BLR size, and accretion state must 
be interpreted as a coupled system.

\subsection{Implications for high-redshift continuum reverberation mapping}

These observations extend continuum reverberation mapping into the sparsely explored regime of luminous quasars in the early Universe.
Despite their extreme luminosities and typically lower variability amplitudes, both objects yield measurable inter-band delays and physically plausible disk scales.
J0529, in particular, demonstrates that continuum lags can be recovered even in the low-amplitude variability regime.

For our fiducial parameters, both quasars lie near nominal thin-disk expectations.
J0529 is consistent with a near-unity disk normalization, while J0920 shows only a marginal excess. 
Notably, neither source requires the factor of $\sim2$--$3$ disk inflation often reported in lower-redshift studies \citep[e.g.][]{Fausnaugh2016,Buitrago25,Mandal25}.
However, our limited wavelength coverage precludes a robust constraint on the lag-wavelength slope ($\beta$).
Our primary result is therefore a constraint on the overall UV-emitting scale rather than a test of departures from the canonical $\lambda^{4/3}$ dependence.

A key strength of this study is the structural context provided by GRAVITY, which reveals a clear radial hierarchy: the UV-emitting disk occupies the innermost few light-days, while the Balmer-line BLR extends to radii two to three orders of magnitude larger. 
Extending this approach to larger samples will be essential for determining whether the consistency with thin-disk theory found here is a universal trait of luminous high-redshift quasars.
In this context, current large time-domain surveys such as the Vera C. Rubin Observatory’s LSST are invaluable for identifying variable quasar populations and providing long-term baseline monitoring (e.g., \citealt{Panda2026} and references therein).
In practice, dedicated follow-up observations remain crucial for isolating continuum-dominated bands and for obtaining the cadence and wavelength coverage required for robust disk-lag measurements. At the same time, the present results suggest that this strategy may be more favorable for high-redshift quasars than for nearby sources, because ground-based optical monitoring samples the rest-frame ultraviolet, where contamination from diffuse BLR continuum emission is expected to be weaker. In this regime, the main requirement is to avoid the strongest UV emission lines, whereas low-redshift monitoring often probes longer rest-frame wavelengths, where the diffuse BLR continuum rises toward the Balmer edge and can remain important even beyond it. High-redshift quasars may therefore offer comparatively cleaner conditions for isolating the intrinsic accretion-disk continuum.

\section{Summary and conclusions}
\label{sum:conclusions}

We present the first continuum reverberation constraints on accretion disk sizes in $z>2$ quasars whose broad-line regions have been spatially resolved by interferometry.
Using medium-band photometric monitoring, we measured inter-band continuum lags in two luminous ($L_{\rm bol} \sim 
10^{48}$\,erg\,s$^{-1}$) quasars: SMSS\,J052915.80$-$435152.0 at $z=3.962$ and SDSS\,J092034.17$+$065718.0 at $z=2.325$.
Both objects have resolved BLRs and dynamical black-hole mass measurements from GRAVITY, making a direct comparison between accretion disk and BLR scales uniquely possible.
Our main results are as follows:

\begin{itemize}
    \item In both sources, we detect statistically significant inter-band continuum lags that increase monotonically with wavelength, as expected for a radially stratified accretion disk.

    \item Modeling the lag spectrum with a fixed thin-disk slope ($\beta=4/3$) yields characteristic UV disk sizes of $R_{\rm AD}=4.35^{+0.78}_{-0.91}$ light-days for J0529 and $R_{\rm AD}=3.15^{+0.50}_{-0.48}$ light-days for J0920. Under the adopted source parameters, both measurements are broadly consistent with nominal thin-disk expectations, with J0920 showing at most a mild excess despite its extreme inferred accretion state.

    \item The two quasars nevertheless illustrate different physical interpretations. For J0529, the inferred disk size agrees with thin-disk expectations if the larger single-epoch black-hole mass is adopted, whereas the smaller GRAVITY+ dynamical mass would imply a more strongly "inflated" disk and a super-Eddington accretion regime. For J0920, conversely, 
the measured disk size independently supports the GRAVITY dynamical mass: the 
larger single-epoch estimates would require a UV continuum region a factor 
of $\sim2$ smaller than thin-disk theory predicts. This highlights the tension between different mass estimates and shows that continuum reverberation measurements provide an independent consistency check on the adopted global source parameters.

    \item J0529 exhibits intrinsic variability at the sub-percent level ($F_{\rm var}\lesssim0.5\%$), yet still yields a measurable lag spectrum. This demonstrates that continuum reverberation mapping remains feasible even for some of the most luminous and weakly variable quasars known, extending the practical reach of the method toward larger and more distant samples.

    \item Comparing the UV disk sizes with the interferometric BLR sizes reveals a pronounced radial hierarchy, with $R_{\rm BLR}/R_{\rm AD}\sim270$ for J0529 (H$\beta$) and $\sim115$ for J0920 (H$\alpha$). This joint analysis links the thermal continuum source to the resolved BLR structure in the same objects, providing a direct empirical map of the inner geometry.
\end{itemize}

Overall, these results demonstrate that continuum reverberation mapping can provide robust empirical constraints on the UV accretion-disk structure of luminous high-redshift quasars. 
While the measured disk scales are broadly consistent with thin-disk theory, their interpretation is coupled to the  black-hole mass, accretion rate, and BLR geometry.
Although based on only two objects, our results also suggest that high-redshift CRM probes shorter rest-frame wavelengths, where the diffuse BLR continuum contributes less; the good agreement with thin-disk predictions found here may partly reflect this weaker contamination, making high-redshift quasars a potentially cleaner probe of disk structure than lower-redshift systems. 
Confirming this --- and disentangling the mass, accretion-rate, and geometry dependencies --- will require combining wide-field time-domain surveys (e.g., Rubin/LSST) with dedicated, finely sampled multi-band follow-up of individual targets.

\begin{acknowledgements}
F. Pozo Nu\~nez gratefully acknowledges the generous and invaluable support of the Klaus Tschira Foundation.
F. Pozo Nu\~nez, A. Mandal, and B. Czerny acknowledge funding from the European Research Council (ERC) under the European Union's Horizon 2020 research and innovation program (grant agreement No 951549).
S. Panda is supported by the international Gemini Observatory, a program of NSF NOIRLab, which is managed by the Association of Universities for Research in Astronomy (AURA) under a cooperative agreement with the U.S. National Science Foundation, on behalf of the Gemini partnership of Argentina, Brazil, Canada, Chile, the Republic of Korea, and the United States of America.
F. Pozo Nu\~nez thanks Christian Wolf and Zachary Steyn for helpful discussions and valuable advice.
We thank the observers and support staff of the MPG/ESO 2.2m telescope at La Silla for carrying out the observations of this long-term monitoring program. We are especially grateful to Maren Hempel and Angela Hempel for their dedicated support of our program and for ensuring that the observations ran smoothly.
We thank the anonymous referee for a careful and constructive report that helped improve the clarity of this paper.
This research has made use of the NASA/IPAC Extragalactic Database (NED) which is operated by the Jet Propulsion Laboratory, California Institute of Technology, under contract with the National Aeronautics and Space Administration. 
This research has made use of the SIMBAD database, operated at CDS, Strasbourg, France.  
      
\end{acknowledgements}

\bibliographystyle{aa}
\bibliography{sn-bibliography}

@ARTICLE{Drewes2026,
       author = {{Drewes}, Farin and {Vieliute}, Roberta and {Hern{\'a}ndez Santisteban}, Juan V. and {Horne}, Keith and {Barth}, Aaron J. and {Cackett}, Edward M. and {Romero Colmenero}, Encarni and {Goad}, Michael R. and {Kaspi}, Shai and {Landt}, Hermine and {Lira}, Paulina and {Netzer}, Hagai and {Vestergaard}, Marianne and {Winkler}, Hartmut},
        title = "{A phenomenological study of the accretion disc in the super-Eddington AGN I Zw 1}",
      journal = {\mnras},
         year = 2026,
        month = mar,
       volume = {546},
       number = {3},
          eid = {stag067},
        pages = {stag067},
          doi = {10.1093/mnras/stag067},
archivePrefix = {arXiv},
       eprint = {2601.05818},
 primaryClass = {astro-ph.GA},
       adsurl = {https://ui.adsabs.harvard.edu/abs/2026MNRAS.546ag067D}
}

@ARTICLE{Lewin2023,
       author = {{Lewin}, Collin and {Kara}, Erin and {Cackett}, Edward M. and {Wilkins}, Dan and {Panagiotou}, Christos and {Garc{\'\i}a}, Javier A. and {Gelbord}, Jonathan},
        title = "{X-Ray/UVOIR Frequency-resolved Time Lag Analysis of Mrk 335 Reveals Accretion Disk Reprocessing}",
      journal = {\apj},
         year = 2023,
        month = sep,
       volume = {954},
       number = {1},
          eid = {33},
        pages = {33},
          doi = {10.3847/1538-4357/ace77b},
archivePrefix = {arXiv},
       eprint = {2307.11145},
 primaryClass = {astro-ph.HE},
       adsurl = {https://ui.adsabs.harvard.edu/abs/2023ApJ...954...33L}
}

@ARTICLE{Cackett2022,
       author = {{Cackett}, Edward M. and {Zoghbi}, Abderahmen and {Ulrich}, Otho},
        title = "{Frequency-resolved Lags in UV/Optical Continuum Reverberation Mapping}",
      journal = {\apj},
         year = 2022,
        month = jan,
       volume = {925},
       number = {1},
          eid = {29},
        pages = {29},
          doi = {10.3847/1538-4357/ac3913},
archivePrefix = {arXiv},
       eprint = {2109.02155},
 primaryClass = {astro-ph.GA},
       adsurl = {https://ui.adsabs.harvard.edu/abs/2022ApJ...925...29C}
}

@ARTICLE{Cackett2020,
       author = {{Cackett}, Edward M. and {Gelbord}, Jonathan and {Li}, Yan-Rong and {Horne}, Keith and {Wang}, Jian-Min and {Barth}, Aaron J. and {Bai}, Jin-Ming and {Bian}, Wei-Hao and {Carroll}, Russell W. and {Du}, Pu and {Edelson}, Rick and {Goad}, Michael R. and {Ho}, Luis C. and {Hu}, Chen and {Khatu}, Viraja C. and {Luo}, Bin and {Miller}, Jake and {Yuan}, Ye-Fei},
        title = "{Supermassive Black Holes with High Accretion Rates in Active Galactic Nuclei. XI. Accretion Disk Reverberation Mapping of Mrk 142}",
      journal = {\apj},
         year = 2020,
        month = jun,
       volume = {896},
       number = {1},
          eid = {1},
        pages = {1},
          doi = {10.3847/1538-4357/ab91b5},
archivePrefix = {arXiv},
       eprint = {2005.03685},
 primaryClass = {astro-ph.HE},
       adsurl = {https://ui.adsabs.harvard.edu/abs/2020ApJ...896....1C}
}

@ARTICLE{korista2001,
       author = {{Korista}, Kirk T. and {Goad}, Michael R.},
        title = "{The Variable Diffuse Continuum Emission of Broad-Line Clouds}",
      journal = {\apj},
         year = 2001,
        month = jun,
       volume = {553},
       number = {2},
        pages = {695-708},
          doi = {10.1086/320964},
archivePrefix = {arXiv},
       eprint = {astro-ph/0101117},
 primaryClass = {astro-ph},
       adsurl = {https://ui.adsabs.harvard.edu/abs/2001ApJ...553..695K}
}

@ARTICLE{jaiswal2025,
       author = {{Jaiswal}, Vikram Kumar and {Mandal}, Amit Kumar and {Prince}, Raj and {Pandey}, Ashwani and {Naddaf}, Mohammad Hassan and {Czerny}, Bo{\.z}ena and {Panda}, Swayamtrupta and {Pozo Nu{\~n}ez}, Francisco},
        title = "{Application of the FRADO model of broad line region formation to Seyfert galaxy NGC 5548 and a first step toward determining the Hubble constant}",
      journal = {\aap},
         year = 2025,
        month = oct,
       volume = {702},
          eid = {A92},
        pages = {A92},
          doi = {10.1051/0004-6361/202452497},
archivePrefix = {arXiv},
       eprint = {2410.03597},
 primaryClass = {astro-ph.GA},
       adsurl = {https://ui.adsabs.harvard.edu/abs/2025A&A...702A..92J}
}

@ARTICLE{2021ApJS..255...20A,
       author = {{Abbott}, T.~M.~C. and {Adam{\'o}w}, M. and {Aguena}, M. and {Allam}, S. and {Amon}, A. and {Annis}, J. and {Avila}, S. and {Bacon}, D. and {Banerji}, M. and {Bechtol}, K. and {Becker}, M.~R. and {Bernstein}, G.~M. and {Bertin}, E. and {Bhargava}, S. and {Bridle}, S.~L. and {Brooks}, D. and {Burke}, D.~L. and {Carnero Rosell}, A. and {Carrasco Kind}, M. and {Carretero}, J. and {Castander}, F.~J. and {Cawthon}, R. and {Chang}, C. and {Choi}, A. and {Conselice}, C. and {Costanzi}, M. and {Crocce}, M. and {da Costa}, L.~N. and {Davis}, T.~M. and {De Vicente}, J. and {DeRose}, J. and {Desai}, S. and {Diehl}, H.~T. and {Dietrich}, J.~P. and {Drlica-Wagner}, A. and {Eckert}, K. and {Elvin-Poole}, J. and {Everett}, S. and {Evrard}, A.~E. and {Ferrero}, I. and {Fert{\'e}}, A. and {Flaugher}, B. and {Fosalba}, P. and {Friedel}, D. and {Frieman}, J. and {Garc{\'\i}a-Bellido}, J. and {Gaztanaga}, E. and {Gelman}, L. and {Gerdes}, D.~W. and {Giannantonio}, T. and {Gill}, M.~S.~S. and {Gruen}, D. and {Gruendl}, R.~A. and {Gschwend}, J. and {Gutierrez}, G. and {Hartley}, W.~G. and {Hinton}, S.~R. and {Hollowood}, D.~L. and {Honscheid}, K. and {Huterer}, D. and {James}, D.~J. and {Jeltema}, T. and {Johnson}, M.~D. and {Kent}, S. and {Kron}, R. and {Kuehn}, K. and {Kuropatkin}, N. and {Lahav}, O. and {Li}, T.~S. and {Lidman}, C. and {Lin}, H. and {MacCrann}, N. and {Maia}, M.~A.~G. and {Manning}, T.~A. and {Maloney}, J.~D. and {March}, M. and {Marshall}, J.~L. and {Martini}, P. and {Melchior}, P. and {Menanteau}, F. and {Miquel}, R. and {Morgan}, R. and {Myles}, J. and {Neilsen}, E. and {Ogando}, R.~L.~C. and {Palmese}, A. and {Paz-Chinch{\'o}n}, F. and {Petravick}, D. and {Pieres}, A. and {Plazas}, A.~A. and {Pond}, C. and {Rodriguez-Monroy}, M. and {Romer}, A.~K. and {Roodman}, A. and {Rykoff}, E.~S. and {Sako}, M. and {Sanchez}, E. and {Santiago}, B. and {Scarpine}, V. and {Serrano}, S. and {Sevilla-Noarbe}, I. and {Smith}, J. Allyn and {Smith}, M. and {Soares-Santos}, M. and {Suchyta}, E. and {Swanson}, M.~E.~C. and {Tarle}, G. and {Thomas}, D. and {To}, C. and {Tremblay}, P.~E. and {Troxel}, M.~A. and {Tucker}, D.~L. and {Turner}, D.~J. and {Varga}, T.~N. and {Walker}, A.~R. and {Wechsler}, R.~H. and {Weller}, J. and {Wester}, W. and {Wilkinson}, R.~D. and {Yanny}, B. and {Zhang}, Y. and {Nikutta}, R. and {Fitzpatrick}, M. and {Jacques}, A. and {Scott}, A. and {Olsen}, K. and {Huang}, L. and {Herrera}, D. and {Juneau}, S. and {Nidever}, D. and {Weaver}, B.~A. and {Adean}, C. and {Correia}, V. and {de Freitas}, M. and {Freitas}, F.~N. and {Singulani}, C. and {Vila-Verde}, G. and {Linea Science Server}},
        title = "{The Dark Energy Survey Data Release 2}",
      journal = {\apjs},
         year = 2021,
        month = aug,
       volume = {255},
       number = {2},
          eid = {20},
        pages = {20},
          doi = {10.3847/1538-4365/ac00b3},
archivePrefix = {arXiv},
       eprint = {2101.05765},
 primaryClass = {astro-ph.IM},
       adsurl = {https://ui.adsabs.harvard.edu/abs/2021ApJS..255...20A}
}

@ARTICLE{2021A&A...650A.154P,
       author = {{Panda}, Swayamtrupta},
        title = "{The CaFe project: Optical Fe II and near-infrared Ca II triplet emission in active galaxies: simulated EWs and the co-dependence of cloud size and metal content}",
      journal = {\aap},
         year = 2021,
        month = jun,
       volume = {650},
          eid = {A154},
        pages = {A154},
          doi = {10.1051/0004-6361/202140393},
archivePrefix = {arXiv},
       eprint = {2004.13113},
 primaryClass = {astro-ph.GA},
       adsurl = {https://ui.adsabs.harvard.edu/abs/2021A&A...650A.154P}
}

@ARTICLE{Zhao2012,
       author = {{Zhao}, Gang and {Zhao}, Yong-Heng and {Chu}, Yao-Quan and {Jing}, Yi-Peng and {Deng}, Li-Cai},
        title = "{LAMOST spectral survey {\textemdash} An overview}",
      journal = {Research in Astronomy and Astrophysics},
         year = 2012,
        month = jul,
       volume = {12},
       number = {7},
        pages = {723-734},
          doi = {10.1088/1674-4527/12/7/002},
       adsurl = {https://ui.adsabs.harvard.edu/abs/2012RAA....12..723Z}
}

@ARTICLE{Baade1999,
       author = {{Baade}, D. and {Meisenheimer}, K. and {Iwert}, O. and {Alonso}, J. and {Augusteijn}, T. and {Beletic}, J. and {Bellemann}, H. and {Benesch}, W. and {B{\"o}hm}, A. and {B{\"o}hnhardt}, H. and {Brewer}, J. and {Deiries}, S. and {Delabre}, B. and {Donaldson}, R. and {Dupuy}, C. and {Franke}, P. and {Gerdes}, R. and {Gilliotte}, A. and {Grimm}, B. and {Haddad}, N. and {Hess}, G. and {Ihle}, G. and {Klein}, R. and {Lenzen}, R. and {Lizon}, J. -L. and {Mancini}, D. and {M{\"u}nch}, N. and {Pizarro}, A. and {Prado}, P. and {Rahmer}, G. and {Reyes}, J. and {Richardson}, F. and {Robledo}, E. and {Sanchez}, F. and {Silber}, A. and {Sinclaire}, P. and {Wackermann}, R. and {Zaggia}, S.},
        title = "{The Wide Field Imager at the 2.2-m MPG/ESO telescope: first views with a 67-million-facette eye.}",
      journal = {The Messenger},
         year = 1999,
        month = mar,
       volume = {95},
        pages = {15-16},
       adsurl = {https://ui.adsabs.harvard.edu/abs/1999Msngr..95...15B}
}

@ARTICLE{Schlafly2011,
       author = {{Schlafly}, Edward F. and {Finkbeiner}, Douglas P.},
        title = "{Measuring Reddening with Sloan Digital Sky Survey Stellar Spectra and Recalibrating SFD}",
      journal = {\apj},
         year = 2011,
        month = aug,
       volume = {737},
       number = {2},
          eid = {103},
        pages = {103},
          doi = {10.1088/0004-637X/737/2/103},
archivePrefix = {arXiv},
       eprint = {1012.4804},
 primaryClass = {astro-ph.GA},
       adsurl = {https://ui.adsabs.harvard.edu/abs/2011ApJ...737..103S}
}

@ARTICLE{Collier1998,
       author = {{Collier}, Stefan and {Horne}, Keith and {Wanders}, Ignaz and {Peterson}, Bradley M.},
        title = "{A new direct method for measuring the Hubble constant from reverberating accretion discs in active galaxies}",
      journal = {\mnras},
         year = 1999,
        month = jan,
       volume = {302},
       number = {1},
        pages = {L24-L28},
          doi = {10.1046/j.1365-8711.1999.02250.x},
archivePrefix = {arXiv},
       eprint = {astro-ph/9811278},
 primaryClass = {astro-ph},
       adsurl = {https://ui.adsabs.harvard.edu/abs/1999MNRAS.302L..24C}
}

@ARTICLE{McHardy2014,
       author = {{McHardy}, I.~M. and {Cameron}, D.~T. and {Dwelly}, T. and {Connolly}, S. and {Lira}, P. and {Emmanoulopoulos}, D. and {Gelbord}, J. and {Breedt}, E. and {Arevalo}, P. and {Uttley}, P.},
        title = "{Swift monitoring of NGC 5548: X-ray reprocessing and short-term UV/optical variability}",
      journal = {\mnras},
         year = 2014,
        month = oct,
       volume = {444},
       number = {2},
        pages = {1469-1474},
          doi = {10.1093/mnras/stu1636},
archivePrefix = {arXiv},
       eprint = {1407.6361},
 primaryClass = {astro-ph.HE},
       adsurl = {https://ui.adsabs.harvard.edu/abs/2014MNRAS.444.1469M}
}

@ARTICLE{Cackett2018,
       author = {{Cackett}, Edward M. and {Chiang}, Chia-Ying and {McHardy}, Ian and {Edelson}, Rick and {Goad}, Michael R. and {Horne}, Keith and {Korista}, Kirk T.},
        title = "{Accretion Disk Reverberation with Hubble Space Telescope Observations of NGC 4593: Evidence for Diffuse Continuum Lags}",
      journal = {\apj},
         year = 2018,
        month = apr,
       volume = {857},
       number = {1},
          eid = {53},
        pages = {53},
          doi = {10.3847/1538-4357/aab4f7},
archivePrefix = {arXiv},
       eprint = {1712.04025},
 primaryClass = {astro-ph.HE},
       adsurl = {https://ui.adsabs.harvard.edu/abs/2018ApJ...857...53C}
}

@ARTICLE{Gonzales2023,
       author = {{Gonz{\'a}lez-Buitrago}, D.~H. and {Garc{\'\i}a-D{\'\i}az}, Ma T. and {Pozo Nu{\~n}ez}, F. and {Guo}, Hengxiao},
        title = "{On the nature of the continuum reverberation of X-ray/UV and optical emission of IRAS 09149-6206}",
      journal = {\mnras},
         year = 2023,
        month = nov,
       volume = {525},
       number = {3},
        pages = {4524-4539},
          doi = {10.1093/mnras/stad2483},
archivePrefix = {arXiv},
       eprint = {2308.05433},
 primaryClass = {astro-ph.HE},
       adsurl = {https://ui.adsabs.harvard.edu/abs/2023MNRAS.525.4524G}
}

@ARTICLE{HernandezSantisteban2020,
       author = {{Hern{\'a}ndez Santisteban}, J.~V. and {Edelson}, R. and {Horne}, K. and {Gelbord}, J.~M. and {Barth}, A.~J. and {Cackett}, E.~M. and {Goad}, M.~R. and {Netzer}, H. and {Starkey}, D. and {Uttley}, P. and {Brandt}, W.~N. and {Korista}, K. and {Lohfink}, A.~M. and {Onken}, C.~A. and {Page}, K.~L. and {Siegel}, M. and {Vestergaard}, M. and {Bisogni}, S. and {Breeveld}, A.~A. and {Cenko}, S.~B. and {Dalla Bont{\`a}}, E. and {Evans}, P.~A. and {Ferland}, G. and {Gonzalez-Buitrago}, D.~H. and {Grupe}, D. and {Joner}, M.~D. and {Kriss}, G. and {LaPorte}, S.~J. and {Mathur}, S. and {Marshall}, F. and {Mehdipour}, M. and {Mudd}, D. and {Peterson}, B.~M. and {Schmidt}, T. and {Vaughan}, S. and {Valenti}, S.},
        title = "{Intensive disc-reverberation mapping of Fairall 9: first year of Swift and LCO monitoring}",
      journal = {\mnras},
         year = 2020,
        month = nov,
       volume = {498},
       number = {4},
        pages = {5399-5416},
          doi = {10.1093/mnras/staa2365},
archivePrefix = {arXiv},
       eprint = {2008.02134},
 primaryClass = {astro-ph.GA},
       adsurl = {https://ui.adsabs.harvard.edu/abs/2020MNRAS.498.5399H}
}

@ARTICLE{Cackett2007,
       author = {{Cackett}, Edward M. and {Horne}, Keith and {Winkler}, Hartmut},
        title = "{Testing thermal reprocessing in active galactic nuclei accretion discs}",
      journal = {\mnras},
         year = 2007,
        month = sep,
       volume = {380},
       number = {2},
        pages = {669-682},
          doi = {10.1111/j.1365-2966.2007.12098.x},
archivePrefix = {arXiv},
       eprint = {0706.1464},
 primaryClass = {astro-ph},
       adsurl = {https://ui.adsabs.harvard.edu/abs/2007MNRAS.380..669C}
}

@ARTICLE{Edelson2019,
       author = {{Edelson}, R. and {Gelbord}, J. and {Cackett}, E. and {Peterson}, B.~M. and {Horne}, K. and {Barth}, A.~J. and {Starkey}, D.~A. and {Bentz}, M. and {Brandt}, W.~N. and {Goad}, M. and {Joner}, M. and {Korista}, K. and {Netzer}, H. and {Page}, K. and {Uttley}, P. and {Vaughan}, S. and {Breeveld}, A. and {Cenko}, S.~B. and {Done}, C. and {Evans}, P. and {Fausnaugh}, M. and {Ferland}, G. and {Gonzalez-Buitrago}, D. and {Gropp}, J. and {Grupe}, D. and {Kaastra}, J. and {Kennea}, J. and {Kriss}, G. and {Mathur}, S. and {Mehdipour}, M. and {Mudd}, D. and {Nousek}, J. and {Schmidt}, T. and {Vestergaard}, M. and {Villforth}, C.},
        title = "{The First Swift Intensive AGN Accretion Disk Reverberation Mapping Survey}",
      journal = {\apj},
         year = 2019,
        month = jan,
       volume = {870},
       number = {2},
          eid = {123},
        pages = {123},
          doi = {10.3847/1538-4357/aaf3b4},
archivePrefix = {arXiv},
       eprint = {1811.07956},
 primaryClass = {astro-ph.HE},
       adsurl = {https://ui.adsabs.harvard.edu/abs/2019ApJ...870..123E}
}

@ARTICLE{Buitrago25,
       author = {{Gonzalez-Buitrago}, D. and {Barth}, A.~J. and {Edelson}, R. and {Hern{\'a}ndez Santisteban}, J.~V. and {Horne}, Keith and {Schmidt}, T. and {Li}, Yan-Rong and {Guo}, Hengxiao and {Joner}, M.~D. and {Cackett}, E. and {Gelbord}, J. and {Bentz}, M.~C. and {Brandt}, W.~N. and {Goad}, M. and {Korista}, K. and {Vestergaard}, M. and {Villforth}, C. and {Breeveld}, A. and {Brink}, T.~G. and {Corsini}, E.~M. and {Dalla Bont{\`a}}, E. and {Ferland}, Gary J. and {Filippenko}, A.~V. and {Garc{\'\i}a-D{\'\i}az}, Ma T. and {Hallum}, M. and {Horst}, J.~C. and {Kim}, M. and {Krongold}, Y. and {Kruger}, J. and {Kuhn}, B. and {Kumar}, S. and {Mehdipour}, M. and {Morelli}, L. and {Mathur}, S. and {Netzer}, H. and {Ochner}, P. and {Pagotto}, I. and {Pizzella}, A. and {Sand}, D.~J. and {Siviero}, A. and {Spencer}, M. and {Sung}, H. and {Vaughan}, S. and {Winkler}, H. and {Zheng}, W.},
        title = "{Departures from standard disc predictions in intensive ground-based monitoring of three AGNs}",
      journal = {\mnras},
         year = 2025,
        month = sep,
       volume = {542},
       number = {3},
        pages = {2572-2596},
          doi = {10.1093/mnras/staf1334},
archivePrefix = {arXiv},
       eprint = {2508.08720},
 primaryClass = {astro-ph.GA},
       adsurl = {https://ui.adsabs.harvard.edu/abs/2025MNRAS.542.2572G}
}

@ARTICLE{Thorne25,
       author = {{Thorne}, James P. and {Landt}, Hermine and {Huang}, Jiamu and {Hern{\'a}ndez Santisteban}, Juan V. and {Horne}, Keith and {Cackett}, Edward M. and {Winkler}, Hartmut and {Sanmartim}, David},
        title = "{Accretion disc reverberation mapping of the quasar 3C 273}",
      journal = {\mnras},
         year = 2025,
        month = mar,
       volume = {537},
       number = {4},
        pages = {3746-3768},
          doi = {10.1093/mnras/staf258},
archivePrefix = {arXiv},
       eprint = {2502.08366},
 primaryClass = {astro-ph.GA},
       adsurl = {https://ui.adsabs.harvard.edu/abs/2025MNRAS.537.3746T}
}

@ARTICLE{Steyn26,
       author = {{Steyn}, Zachary and {Wolf}, Christian and {Onken}, Christopher and {Smith}, Ken and {Tang}, Ji-Jia and {Kova{\v{c}}evi{\'c}}, Andjelka B. and {Tonry}, John and {Clocchiatti}, Alejandro},
        title = "{Continuum reverberation in bright quasars using NASA/ATLAS}",
      journal = {\mnras},
         year = 2026,
        month = jun,
       volume = {548},
       number = {4},
          eid = {stag702},
        pages = {stag702},
          doi = {10.1093/mnras/stag702},
archivePrefix = {arXiv},
       eprint = {2603.11789},
 primaryClass = {astro-ph.GA},
       adsurl = {https://ui.adsabs.harvard.edu/abs/2026MNRAS.548ag702S}
}

@ARTICLE{Leung2026,
       author = {{Leung}, Gene C.~K. and {Eilers}, Anna-Christina and {Panagiotou}, Christos and {Wolf}, Julien and {De}, Kishalay and {Weisenbach}, Luke and {Yue}, Minghao and {Fan}, Xiaohui and {Ishikawa}, Yuzo and {Kara}, Erin and {Krumpe}, Mirko and {Merloni}, Andrea and {Simcoe}, Robert A. and {Wang}, Feige and {Yang}, Jinyi},
        title = "{Discovery of quasar variability and early accretion disk signatures at cosmic dawn}",
      journal = {Nature Astronomy},
         year = 2026,
        month = jun,
          doi = {10.1038/s41550-026-02897-4},
archivePrefix = {arXiv},
       eprint = {2605.00978},
 primaryClass = {astro-ph.GA},
       adsurl = {https://ui.adsabs.harvard.edu/abs/2026NatAs.tmp..123L}
}

@ARTICLE{Bentz2010,
       author = {{Bentz}, Misty C. and {Walsh}, Jonelle L. and {Barth}, Aaron J. and {Yoshii}, Yuzuru and {Woo}, Jong-Hak and {Wang}, Xiaofeng and {Treu}, Tommaso and {Thornton}, Carol E. and {Street}, Rachel A. and {Steele}, Thea N. and {Silverman}, Jeffrey M. and {Serduke}, Frank J.~D. and {Sakata}, Yu and {Minezaki}, Takeo and {Malkan}, Matthew A. and {Li}, Weidong and {Lee}, Nicholas and {Hiner}, Kyle D. and {Hidas}, Marton G. and {Greene}, Jenny E. and {Gates}, Elinor L. and {Ganeshalingam}, Mohan and {Filippenko}, Alexei V. and {Canalizo}, Gabriela and {Bennert}, Vardha Nicola and {Baliber}, Nairn},
        title = "{The Lick AGN Monitoring Project: Reverberation Mapping of Optical Hydrogen and Helium Recombination Lines}",
      journal = {\apj},
         year = 2010,
        month = jun,
       volume = {716},
       number = {2},
        pages = {993-1011},
          doi = {10.1088/0004-637X/716/2/993},
archivePrefix = {arXiv},
       eprint = {1004.2922},
 primaryClass = {astro-ph.CO},
       adsurl = {https://ui.adsabs.harvard.edu/abs/2010ApJ...716..993B}
}

@ARTICLE{McDougall2025,
       author = {{McDougall}, Hugh Gareth and {Davis}, Tamara M. and {Yu}, Zhefu and {Martini}, Paul and {Lidman}, Chris and {Malik}, Umang and {Penton}, Andrew and {Lewis}, Geraint and {Tucker}, Brad E. and {Pope}, Benjamin and {Allam}, Sahar and {Andrade-Oliveira}, Felipe and {Asorey}, Jacobo and {Bacon}, David and {Bocquet}, Sebastian and {Brooks}, David and {Carnero Rosell}, Aurelio and {Carollo}, Daniela and {Carr}, Anthony and {Carretero}, Jorge and {Cheng}, Ting-Yun and {Da Costa}, Luiz and {da Silva Pereira}, Maria Elidaiana and {De Vicente}, Juan and {Thomas Diehl}, H. and {Doel}, Peter and {Everett}, Spencer and {Garcia-Bellido}, Juan and {Glazebrook}, Karl and {Gruen}, Daniel and {Gutierrez}, Gaston and {Herner}, Kenneth and {Hinton}, Samuel R. and {Hollowood}, Daniel and {James}, David and {Kim}, Alex and {Kuehn}, Kyler and {Lee}, Sujeong and {March}, Marisa and {Marshall}, Jennifer and {Mena-Fernandez}, Juan and {Miquel}, Ramon and {Myles}, Justin and {Nichol}, Robert and {Ogando}, Ricardo and {Porredon}, Anna and {Sanchez}, Eusebio and {Sanchez Cid}, David and {Sharp}, Rob and {Smith}, Mathew and {Suchyta}, Eric and {Swanson}, Molly and {To}, Chun-Hao and {Tucker}, Douglas and {Walker}, Alistair and {Weaverdyck}, Noah},
        title = "{OzDES reverberation mapping of Active Galactic Nuclei: Final data release, black-hole mass results, and scaling relations}",
      journal = {\pasa},
         year = 2026,
        month = jun,
       volume = {43},
          eid = {e084},
        pages = {e084},
          doi = {10.1017/pasa.2026.10219},
archivePrefix = {arXiv},
       eprint = {2512.01261},
 primaryClass = {astro-ph.GA},
       adsurl = {https://ui.adsabs.harvard.edu/abs/2026PASA...43...84M}
}

@ARTICLE{Mandal25,
       author = {{Mandal}, Amit Kumar and {Woo}, Jong-Hak and {Wang}, Shu},
        title = "{The Size of the Continuum Emission Region and Its Scaling Relations with Active Galactic Nucleus Luminosity and the Broad-line Region Size}",
      journal = {\apj},
         year = 2025,
        month = may,
       volume = {985},
       number = {1},
          eid = {30},
        pages = {30},
          doi = {10.3847/1538-4357/adc56e},
archivePrefix = {arXiv},
       eprint = {2502.19184},
 primaryClass = {astro-ph.GA},
       adsurl = {https://ui.adsabs.harvard.edu/abs/2025ApJ...985...30M}
}

@ARTICLE{Fausnaugh2016,
       author = {{Fausnaugh}, M.~M. and {Denney}, K.~D. and {Barth}, A.~J. and {Bentz}, M.~C. and {Bottorff}, M.~C. and {Carini}, M.~T. and {Croxall}, K.~V. and {De Rosa}, G. and {Goad}, M.~R. and {Horne}, Keith and {Joner}, M.~D. and {Kaspi}, S. and {Kim}, M. and {Klimanov}, S.~A. and {Kochanek}, C.~S. and {Leonard}, D.~C. and {Netzer}, H. and {Peterson}, B.~M. and {Schn{\"u}lle}, K. and {Sergeev}, S.~G. and {Vestergaard}, M. and {Zheng}, W.-K. and {Zu}, Y. and {Anderson}, M.~D. and {Ar{\'e}valo}, P. and {Bazhaw}, C. and {Borman}, G.~A. and {Boroson}, T.~A. and {Brandt}, W.~N. and {Breeveld}, A.~A. and {Brewer}, B.~J. and {Cackett}, E.~M. and {Crenshaw}, D.~M. and {Dalla Bont{\`a}}, E. and {De Lorenzo-C{\'a}ceres}, A. and {Dietrich}, M. and {Edelson}, R. and {Efimova}, N.~V. and {Ely}, J. and {Evans}, P.~A. and {Filippenko}, A.~V. and {Flatland}, K. and {Gehrels}, N. and {Geier}, S. and {Gelbord}, J.~M. and {Gonzalez}, L. and {Gorjian}, V. and {Grier}, C.~J. and {Grupe}, D. and {Hall}, P.~B. and {Hicks}, S. and {Horenstein}, D. and {Hutchison}, T. and {Im}, M. and {Jensen}, J.~J. and {Jones}, J. and {Kaastra}, J. and {Kelly}, B.~C. and {Kennea}, J.~A. and {Kim}, S.~C. and {Korista}, K.~T. and {Kriss}, G.~A. and {Lee}, J.~C. and {Lira}, P. and {MacInnis}, F. and {Manne-Nicholas}, E.~R. and {Mathur}, S. and {McHardy}, I.~M. and {Montouri}, C. and {Musso}, R. and {Nazarov}, S.~V. and {Norris}, R.~P. and {Nousek}, J.~A. and {Okhmat}, D.~N. and {Pancoast}, A. and {Papadakis}, I. and {Parks}, J.~R. and {Pei}, L. and {Pogge}, R.~W. and {Pott}, J.-U. and {Rafter}, S.~E. and {Rix}, H.-W. and {Saylor}, D.~A. and {Schimoia}, J.~S. and {Siegel}, M. and {Spencer}, M. and {Starkey}, D. and {Sung}, H.-I. and {Teems}, K.~G. and {Treu}, T. and {Turner}, C.~S. and {Uttley}, P. and {Villforth}, C. and {Weiss}, Y. and {Woo}, J.-H. and {Yan}, H. and {Young}, S.},
        title = "{Space Telescope and Optical Reverberation Mapping Project. III. Optical Continuum Emission and Broadband Time Delays in NGC 5548}",
      journal = {\apj},
         year = 2016,
        month = apr,
       volume = {821},
       number = {1},
          eid = {56},
        pages = {56},
          doi = {10.3847/0004-637X/821/1/56},
archivePrefix = {arXiv},
       eprint = {1510.05648},
 primaryClass = {astro-ph.GA},
       adsurl = {https://ui.adsabs.harvard.edu/abs/2016ApJ...821...56F}
}

@ARTICLE{Edelson2017,
       author = {{Edelson}, R. and {Gelbord}, J. and {Cackett}, E. and {Connolly}, S. and {Done}, C. and {Fausnaugh}, M. and {Gardner}, E. and {Gehrels}, N. and {Goad}, M. and {Horne}, K. and {McHardy}, I. and {Peterson}, B.~M. and {Vaughan}, S. and {Vestergaard}, M. and {Breeveld}, A. and {Barth}, A.~J. and {Bentz}, M. and {Bottorff}, M. and {Brandt}, W.~N. and {Crawford}, S.~M. and {Dalla Bont{\`a}}, E. and {Emmanoulopoulos}, D. and {Evans}, P. and {Figuera Jaimes}, R. and {Filippenko}, A.~V. and {Ferland}, G. and {Grupe}, D. and {Joner}, M. and {Kennea}, J. and {Korista}, K.~T. and {Krimm}, H.~A. and {Kriss}, G. and {Leonard}, D.~C. and {Mathur}, S. and {Netzer}, H. and {Nousek}, J. and {Page}, K. and {Romero-Colmenero}, E. and {Siegel}, M. and {Starkey}, D.~A. and {Treu}, T. and {Vogler}, H.~A. and {Winkler}, H. and {Zheng}, W.},
        title = "{Swift Monitoring of NGC 4151: Evidence for a Second X-Ray/UV Reprocessing}",
      journal = {\apj},
         year = 2017,
        month = may,
       volume = {840},
       number = {1},
          eid = {41},
        pages = {41},
          doi = {10.3847/1538-4357/aa6890},
archivePrefix = {arXiv},
       eprint = {1703.06901},
 primaryClass = {astro-ph.HE},
       adsurl = {https://ui.adsabs.harvard.edu/abs/2017ApJ...840...41E}
}

@ARTICLE{Papadakis2022,
       author = {{Papadakis}, I.~E. and {Dov{\v{c}}iak}, M. and {Kammoun}, E.~S.},
        title = "{X-ray illuminated accretion discs and quasar microlensing disc sizes}",
      journal = {\aap},
         year = 2022,
        month = oct,
       volume = {666},
          eid = {A11},
        pages = {A11},
          doi = {10.1051/0004-6361/202142962},
archivePrefix = {arXiv},
       eprint = {2207.12473},
 primaryClass = {astro-ph.HE},
       adsurl = {https://ui.adsabs.harvard.edu/abs/2022A&A...666A..11P}
}

@ARTICLE{Kammoun2023,
       author = {{Kammoun}, E.~S. and {Robin}, L. and {Papadakis}, I.~E. and {Dov{\v{c}}iak}, M. and {Panagiotou}, C.},
        title = "{Revisiting UV/optical continuum time lags in AGN}",
      journal = {\mnras},
         year = 2023,
        month = nov,
       volume = {526},
       number = {1},
        pages = {138-151},
          doi = {10.1093/mnras/stad2701},
archivePrefix = {arXiv},
       eprint = {2309.05392},
 primaryClass = {astro-ph.HE},
       adsurl = {https://ui.adsabs.harvard.edu/abs/2023MNRAS.526..138K}
}

@ARTICLE{Panda2024,
       author = {{Panda}, Swayamtrupta and {Pozo Nu{\~n}ez}, Francisco and {Ba{\~n}ados}, Eduardo and {Heidt}, Jochen},
        title = "{Probing the C IV Continuum Size{\textendash}Luminosity Relation in Active Galactic Nuclei with Photometric Reverberation Mapping}",
      journal = {\apjl},
         year = 2024,
        month = jun,
       volume = {968},
       number = {2},
          eid = {L16},
        pages = {L16},
          doi = {10.3847/2041-8213/ad5014},
archivePrefix = {arXiv},
       eprint = {2405.11649},
 primaryClass = {astro-ph.GA},
       adsurl = {https://ui.adsabs.harvard.edu/abs/2024ApJ...968L..16P}
}

@ARTICLE{Sharp2024,
       author = {{Sharp}, Hugh W. and {Homayouni}, Y. and {Trump}, Jonathan R. and {Anderson}, Scott F. and {Assef}, Roberto J. and {Brandt}, W.~N. and {Davis}, Megan C. and {Fries}, Logan B. and {Grier}, Catherine J. and {Hall}, Patrick B. and {Horne}, Keith and {Koekemoer}, Anton M. and {Mart{\'\i}nez-Aldama}, Mary Loli and {Menezes}, David M. and {Pena}, Theodore and {Ricci}, C. and {Schneider}, Donald P. and {Shen}, Yue and {Trakhtenbrot}, Benny},
        title = "{The Sloan Digital Sky Survey Reverberation Mapping Project: Investigation of Continuum Lag Dependence on Broad-line Contamination and Quasar Properties}",
      journal = {\apj},
         year = 2024,
        month = jan,
       volume = {961},
       number = {1},
          eid = {93},
        pages = {93},
          doi = {10.3847/1538-4357/ad0cea},
archivePrefix = {arXiv},
       eprint = {2309.02499},
 primaryClass = {astro-ph.GA},
       adsurl = {https://ui.adsabs.harvard.edu/abs/2024ApJ...961...93S}
}

@ARTICLE{Homayouni2019,
       author = {{Homayouni}, Y. and {Trump}, Jonathan R. and {Grier}, C.~J. and {Shen}, Yue and {Starkey}, D.~A. and {Brandt}, W.~N. and {Fonseca Alvarez}, G. and {Hall}, P.~B. and {Horne}, Keith and {Kinemuchi}, Karen and {I-Hsiu Li}, Jennifer and {McGreer}, Ian D. and {Sun}, Mouyuan and {Ho}, L.~C. and {Schneider}, D.~P.},
        title = "{The Sloan Digital Sky Survey Reverberation Mapping Project: Accretion Disk Sizes from Continuum Lags}",
      journal = {\apj},
         year = 2019,
        month = aug,
       volume = {880},
       number = {2},
          eid = {126},
        pages = {126},
          doi = {10.3847/1538-4357/ab2638},
archivePrefix = {arXiv},
       eprint = {1806.08360},
 primaryClass = {astro-ph.GA},
       adsurl = {https://ui.adsabs.harvard.edu/abs/2019ApJ...880..126H}
}

@ARTICLE{Sergeev2005,
       author = {{Sergeev}, S.~G. and {Doroshenko}, V.~T. and {Golubinskiy}, Yu. V. and {Merkulova}, N.~I. and {Sergeeva}, E.~A.},
        title = "{Lag-Luminosity Relationship for Interband Lags between Variations in B, V, R, and I Bands in Active Galactic Nuclei}",
      journal = {\apj},
         year = 2005,
        month = mar,
       volume = {622},
       number = {1},
        pages = {129-135},
          doi = {10.1086/427820},
       adsurl = {https://ui.adsabs.harvard.edu/abs/2005ApJ...622..129S}
}

@ARTICLE{2022MNRAS.509.2637N,
       author = {{Netzer}, Hagai},
        title = "{Continuum reverberation mapping and a new lag-luminosity relationship for AGN}",
      journal = {\mnras},
         year = 2022,
        month = jan,
       volume = {509},
       number = {2},
        pages = {2637-2646},
          doi = {10.1093/mnras/stab3133},
archivePrefix = {arXiv},
       eprint = {2110.05512},
 primaryClass = {astro-ph.GA},
       adsurl = {https://ui.adsabs.harvard.edu/abs/2022MNRAS.509.2637N}
}

@incollection{Novikov1973,
  author    = {Novikov, I. D. and Thorne, K. S.},
  title     = {Astrophysics of Black Holes},
  booktitle = {Black Holes (Les Astres Occlus)},
  editor    = {DeWitt, C. and DeWitt, B. S.},
  publisher = {Gordon and Breach},
  address   = {New York},
  year      = {1973},
  pages     = {343--450}
}

@ARTICLE{Shakura1973,
       author = {{Shakura}, N.~I. and {Sunyaev}, R.~A.},
        title = "{Black holes in binary systems. Observational appearance.}",
      journal = {\aap},
         year = 1973,
        month = jan,
       volume = {24},
        pages = {337-355},
       adsurl = {https://ui.adsabs.harvard.edu/abs/1973A&A....24..337S}
}

@ARTICLE{Gaskell2017,
       author = {{Gaskell}, C. Martin},
        title = "{The case for cases B and C: intrinsic hydrogen line ratios of the broad-line region of active galactic nuclei, reddenings, and accretion disc sizes}",
      journal = {\mnras},
         year = 2017,
        month = may,
       volume = {467},
       number = {1},
        pages = {226-238},
          doi = {10.1093/mnras/stx094},
archivePrefix = {arXiv},
       eprint = {1512.09291},
 primaryClass = {astro-ph.GA},
       adsurl = {https://ui.adsabs.harvard.edu/abs/2017MNRAS.467..226G}
}

@ARTICLE{Dexter2011,
       author = {{Dexter}, Jason and {Agol}, Eric},
        title = "{Quasar Accretion Disks are Strongly Inhomogeneous}",
      journal = {\apjl},
         year = 2011,
        month = jan,
       volume = {727},
       number = {1},
          eid = {L24},
        pages = {L24},
          doi = {10.1088/2041-8205/727/1/L24},
archivePrefix = {arXiv},
       eprint = {1012.3169},
 primaryClass = {astro-ph.CO},
       adsurl = {https://ui.adsabs.harvard.edu/abs/2011ApJ...727L..24D}
}

@ARTICLE{Hall2018,
       author = {{Hall}, Patrick B. and {Sarrouh}, Ghassan T. and {Horne}, Keith},
        title = "{Non-blackbody Disks Can Help Explain Inferred AGN Accretion Disk Sizes}",
      journal = {\apj},
         year = 2018,
        month = feb,
       volume = {854},
       number = {2},
          eid = {93},
        pages = {93},
          doi = {10.3847/1538-4357/aaa768},
archivePrefix = {arXiv},
       eprint = {1705.05467},
 primaryClass = {astro-ph.GA},
       adsurl = {https://ui.adsabs.harvard.edu/abs/2018ApJ...854...93H}
}

@ARTICLE{Weaver2022,
       author = {{Weaver}, John R. and {Horne}, Keith},
        title = "{Dust and the intrinsic spectral index of quasar variations: hints of finite stress at the innermost stable circular orbit}",
      journal = {\mnras},
         year = 2022,
        month = may,
       volume = {512},
       number = {1},
        pages = {899-916},
          doi = {10.1093/mnras/stac248},
archivePrefix = {arXiv},
       eprint = {2201.11134},
 primaryClass = {astro-ph.GA},
       adsurl = {https://ui.adsabs.harvard.edu/abs/2022MNRAS.512..899W}
}

@ARTICLE{1997ApJS..110....9R,
       author = {{Rodr{\'\i}guez-Pascual}, P.~M. and {Alloin}, D. and {Clavel}, J. and {Crenshaw}, D.~M. and {Horne}, K. and {Kriss}, G.~A. and {Krolik}, J.~H. and {Malkan}, M.~A. and {Netzer}, H. and {O'Brien}, P.~T. and {Peterson}, B.~M. and {Reichert}, G.~A. and {Wamsteker}, W. and {Alexander}, T. and {Barr}, P. and {Blandford}, R.~D. and {Bregman}, J.~N. and {Carone}, T.~E. and {Clements}, S. and {Courvoisier}, T.-J. and {De Robertis}, M.~M. and {Dietrich}, M. and {Dottori}, H. and {Edelson}, R.~A. and {Filippenko}, A.~V. and {Gaskell}, C.~M. and {Huchra}, J.~P. and {Hutchings}, J.~B. and {Kollatschny}, W. and {Koratkar}, A.~P. and {Korista}, K.~T. and {Laor}, A. and {MacAlpine}, G.~M. and {Martin}, P.~G. and {Maoz}, D. and {McCollum}, B. and {Morris}, S.~L. and {Perola}, G.~C. and {Pogge}, R.~W. and {Ptak}, R.~L. and {Recondo-Gonz{\'a}lez}, M.~C. and {Rodr{\'\i}guez-Espinoza}, J.~M. and {Rokaki}, E.~L. and {Santos-Lle{\'o}}, M. and {Sekiguchi}, K. and {Shull}, J.~M. and {Snijders}, M.~A.~J. and {Sparke}, L.~S. and {Stirpe}, G.~M. and {Stoner}, R.~E. and {Sun}, W.-H. and {Wagner}, S.~J. and {Wanders}, I. and {Wilkes}, J. and {Winge}, C. and {Zheng}, W.},
        title = "{Steps toward Determination of the Size and Structure of the Broad-Line Region in Active Galactic Nuclei. IX. Ultraviolet Observations of Fairall 9}",
      journal = {\apjs},
         year = 1997,
        month = may,
       volume = {110},
       number = {1},
        pages = {9-20},
          doi = {10.1086/312996},
       adsurl = {https://ui.adsabs.harvard.edu/abs/1997ApJS..110....9R}
}

@ARTICLE{2024ApJ...962...67W,
       author = {{Woo}, Jong-Hak and {Wang}, Shu and {Rakshit}, Suvendu and {Cho}, Hojin and {Son}, Donghoon and {Bennert}, Vardha N. and {Gallo}, Elena and {Hodges-Kluck}, Edmund and {Treu}, Tommaso and {Barth}, Aaron J. and {Cho}, Wanjin and {Foord}, Adi and {Geum}, Jaehyuk and {Guo}, Hengxiao and {Jadhav}, Yashashree and {Jeon}, Yiseul and {Kabasares}, Kyle M. and {Kang}, Won-Suk and {Kim}, Changseok and {Kim}, Minjin and {Kim}, Tae-Woo and {Le}, Huynh Anh N. and {Malkan}, Matthew A. and {Mandal}, Amit Kumar and {Park}, Daeseong and {Spencer}, Chance and {Shin}, Jaejin and {Sung}, Hyun-il and {U}, Vivian and {Williams}, Peter R. and {Yee}, Nick},
        title = "{The Seoul National University AGN Monitoring Project. III. H{\ensuremath{\beta}} Lag Measurements of 32 Luminous Active Galactic Nuclei and the High-luminosity End of the Size─Luminosity Relation}",
      journal = {\apj},
         year = 2024,
        month = feb,
       volume = {962},
       number = {1},
          eid = {67},
        pages = {67},
          doi = {10.3847/1538-4357/ad132f},
archivePrefix = {arXiv},
       eprint = {2311.15518},
 primaryClass = {astro-ph.GA},
       adsurl = {https://ui.adsabs.harvard.edu/abs/2024ApJ...962...67W}
}

@ARTICLE{2026A&A...706A.176M,
       author = {{Mandal}, Amit Kumar and {Pozo Nu{\~n}ez}, Francisco and {Jaiswal}, Vikram Kumar and {Naddaf}, Mohammad Hassan and {Czerny}, Bo{\.z}ena and {Panda}, Swayamtrupta and {Karczmarek}, Paulina and {Pietrzy{\'n}ski}, Grzegorz and {Pandey}, Shivangi and {Peterson}, B.~M. and {Zaja{\v{c}}ek}, Michal and {Dov{\v{c}}iak}, Michal and {Karas}, Vladimir and {Narloch}, Weronika and {Kicia}, Miros{\l}aw and {G{\'o}rski}, Marek and {Ka{\l}uszy{\'n}ski}, Miko{\l}aj and {Hajdu}, Gergely and {Wielg{\'o}rski}, Piotr and {Zgirski}, Bart{\l}omiej and {Ga{\l}an}, Cezary and {Pych}, Wojciech and {Smolec}, Rados{\l}aw and {B{\k{a}}kowska}, Karolina and {Gieren}, Wolfgang and {Kervella}, Pierre},
        title = "{HALO: I. Photometric continuum reverberation mapping of Fairall 9}",
      journal = {\aap},
         year = 2026,
        month = feb,
       volume = {706},
          eid = {A176},
        pages = {A176},
          doi = {10.1051/0004-6361/202557795},
archivePrefix = {arXiv},
       eprint = {2512.13296},
 primaryClass = {astro-ph.GA},
       adsurl = {https://ui.adsabs.harvard.edu/abs/2026A&A...706A.176M}
}

@ARTICLE{1998ApJ...500..162C,
       author = {{Collier}, S.~J. and {Horne}, Keith and {Kaspi}, S. and {Netzer}, H. and {Peterson}, B.~M. and {Wanders}, I. and {Alexander}, T. and {Bertram}, R. and {Comastri}, A. and {Gaskell}, C.~M. and {Malkov}, Yu. F. and {Maoz}, D. and {Mignoli}, M. and {Pogge}, R.~W. and {Pronik}, V.~I. and {Sergeev}, S.~G. and {Snedden}, S. and {Stirpe}, G.~M. and {Bochkarev}, N.~G. and {Burenkov}, A.~N. and {Shapovalova}, A.~I. and {Wagner}, R.~M.},
        title = "{Steps toward Determination of the Size and Structure of the Broad-Line Region in Active Galactic Nuclei. XIV. Intensive Optical Spectrophotometric Observations of NGC 7469}",
      journal = {\apj},
         year = 1998,
        month = jun,
       volume = {500},
       number = {1},
        pages = {162-172},
          doi = {10.1086/305720},
       adsurl = {https://ui.adsabs.harvard.edu/abs/1998ApJ...500..162C}
}

@ARTICLE{Bate2008,
       author = {{Bate}, N.~F. and {Floyd}, D.~J.~E. and {Webster}, R.~L. and {Wyithe}, J.~S.~B.},
        title = "{A microlensing study of the accretion disc in the quasar MG 0414+0534}",
      journal = {\mnras},
         year = 2008,
        month = dec,
       volume = {391},
       number = {4},
        pages = {1955-1960},
          doi = {10.1111/j.1365-2966.2008.14020.x},
archivePrefix = {arXiv},
       eprint = {0810.1092},
 primaryClass = {astro-ph},
       adsurl = {https://ui.adsabs.harvard.edu/abs/2008MNRAS.391.1955B}
}

@ARTICLE{Cornachione2020,
       author = {{Cornachione}, Matthew A. and {Morgan}, Christopher W. and {Millon}, Martin and {Bentz}, Misty C. and {Courbin}, Frederic and {Bonvin}, Vivien and {Falco}, Emilio E.},
        title = "{A Microlensing Accretion Disk Size Measurement in the Lensed Quasar WFI 2026-4536}",
      journal = {\apj},
         year = 2020,
        month = jun,
       volume = {895},
       number = {2},
          eid = {125},
        pages = {125},
          doi = {10.3847/1538-4357/ab557a},
archivePrefix = {arXiv},
       eprint = {1911.06218},
 primaryClass = {astro-ph.HE},
       adsurl = {https://ui.adsabs.harvard.edu/abs/2020ApJ...895..125C}
}

@ARTICLE{Marculewicz2024,
       author = {{Marculewicz}, Marcin and {Sun}, Mouyuan and {Zhang}, Zhixiang and {Yi}, Tuan},
        title = "{The Disk Reverberation Mapping of the Lensed Quasar Q0957+561}",
      journal = {\apj},
         year = 2024,
        month = dec,
       volume = {976},
       number = {2},
          eid = {211},
        pages = {211},
          doi = {10.3847/1538-4357/ad8b1a},
archivePrefix = {arXiv},
       eprint = {2410.17549},
 primaryClass = {astro-ph.HE},
       adsurl = {https://ui.adsabs.harvard.edu/abs/2024ApJ...976..211M}
}

@ARTICLE{2019ApJ...886...42D,
       author = {{Du}, Pu and {Wang}, Jian-Min},
        title = "{The Radius-Luminosity Relationship Depends on Optical Spectra in Active Galactic Nuclei}",
      journal = {\apj},
         year = 2019,
        month = nov,
       volume = {886},
       number = {1},
          eid = {42},
        pages = {42},
          doi = {10.3847/1538-4357/ab4908},
archivePrefix = {arXiv},
       eprint = {1909.06735},
 primaryClass = {astro-ph.GA},
       adsurl = {https://ui.adsabs.harvard.edu/abs/2019ApJ...886...42D}
}

@ARTICLE{Gaskell2023,
       author = {{Gaskell}, C. Martin and {Anderson}, Frances C. and {Birmingham}, Sufia {\'A}. and {Ghosh}, Samhita},
        title = "{Estimating reddening of the continuum and broad-line region of active galactic nuclei: the mean reddening of NGC 5548 and the size of the accretion disc}",
      journal = {\mnras},
         year = 2023,
        month = mar,
       volume = {519},
       number = {3},
        pages = {4082-4093},
          doi = {10.1093/mnras/stac3333},
archivePrefix = {arXiv},
       eprint = {2208.11437},
 primaryClass = {astro-ph.GA},
       adsurl = {https://ui.adsabs.harvard.edu/abs/2023MNRAS.519.4082G}
}

@ARTICLE{Sorgenfrei2025,
       author = {{Sorgenfrei}, C. and {Schmidt}, R.~W. and {Wambsganss}, J.},
        title = "{Detection of colour variations from gravitational microlensing observations in the quadruple quasar HE0435-1223: Implications for the accretion disc}",
      journal = {\aap},
         year = 2025,
        month = nov,
       volume = {703},
          eid = {A250},
        pages = {A250},
          doi = {10.1051/0004-6361/202555933},
archivePrefix = {arXiv},
       eprint = {2509.09341},
 primaryClass = {astro-ph.GA},
       adsurl = {https://ui.adsabs.harvard.edu/abs/2025A&A...703A.250S}
}

@ARTICLE{Panda2026,
       author = {{Panda}, Swayamtrupta and {Pozo Nu{\~n}ez}, Francisco and {Benati Gon{\c{c}}alves}, Hygor and {Li}, Guodong and {Czerny}, Bo{\.z}ena and {Marziani}, Paola and {Storchi-Bergmann}, Thaisa},
        title = "{AGN Variability with Rubin Observatory in the 2030s}",
      journal = {arXiv e-prints},
         year = 2026,
        month = jan,
          eid = {arXiv:2601.21769},
        pages = {arXiv:2601.21769},
          doi = {10.48550/arXiv.2601.21769},
archivePrefix = {arXiv},
       eprint = {2601.21769},
 primaryClass = {astro-ph.GA},
       adsurl = {https://ui.adsabs.harvard.edu/abs/2026arXiv260121769P}
}

@ARTICLE{2023ApJ...958..195C,
       author = {{Cackett}, Edward M. and {Gelbord}, Jonathan and {Barth}, Aaron J. and {De Rosa}, Gisella and {Edelson}, Rick and {Goad}, Michael R. and {Homayouni}, Yasaman and {Horne}, Keith and {Kara}, Erin A. and {Kriss}, Gerard A. and {Korista}, Kirk T. and {Landt}, Hermine and {Plesha}, Rachel and {Arav}, Nahum and {Bentz}, Misty C. and {Boizelle}, Benjamin D. and {Dalla Bont{\`a}}, Elena and {Dehghanian}, Maryam and {Donnan}, Fergus and {Du}, Pu and {Ferland}, Gary J. and {Fian}, Carina and {Filippenko}, Alexei V. and {Gonz{\'a}lez Buitrago}, Diego H. and {Grier}, Catherine J. and {Hall}, Patrick B. and {Hu}, Chen and {Ili{\'c}}, Dragana and {Kaastra}, Jelle and {Kaspi}, Shai and {Kochanek}, Christopher S. and {Kova{\v{c}}evi{\'c}}, Andjelka B. and {Kynoch}, Daniel and {Li}, Yan-Rong and {McLane}, Jacob N. and {Mehdipour}, Missagh and {Miller}, Jake A. and {Montano}, John and {Netzer}, Hagai and {Panagiotou}, Christos and {Partington}, Ethan and {{\v{C}}. Popovi{\'c}}, Luka and {Proga}, Daniel and {Rogantini}, Daniele and {Sanmartim}, David and {Siebert}, Matthew R. and {Storchi-Bergmann}, Thaisa and {Vestergaard}, Marianne and {Wang}, Jian-Min and {Waters}, Tim and {Zaidouni}, Fatima},
        title = "{AGN STORM 2. IV. Swift X-Ray and Ultraviolet/Optical Monitoring of Mrk 817}",
      journal = {\apj},
         year = 2023,
        month = dec,
       volume = {958},
       number = {2},
          eid = {195},
        pages = {195},
          doi = {10.3847/1538-4357/acfdac},
archivePrefix = {arXiv},
       eprint = {2306.17663},
 primaryClass = {astro-ph.HE},
       adsurl = {https://ui.adsabs.harvard.edu/abs/2023ApJ...958..195C}
}

@ARTICLE{Tang2024,
       author = {{Tang}, Ji-Jia and {Wolf}, Christian and {Tonry}, John},
        title = "{The variability structure function of the highest luminosity quasars on short time-scales}",
      journal = {\mnras},
         year = 2024,
        month = dec,
       volume = {535},
       number = {3},
        pages = {2260-2268},
          doi = {10.1093/mnras/stae2479},
archivePrefix = {arXiv},
       eprint = {2411.07280},
 primaryClass = {astro-ph.GA},
       adsurl = {https://ui.adsabs.harvard.edu/abs/2024MNRAS.535.2260T}
}

@ARTICLE{Stone2023,
       author = {{Stone}, Zachary and {Shen}, Yue and {Burke}, Colin J. and {Chen}, Yu-Ching and {Yang}, Qian and {Liu}, Xin and {Gruendl}, R.~A. and {Adam{\'o}w}, M. and {Andrade-Oliveira}, F. and {Annis}, J. and {Bacon}, D. and {Bertin}, E. and {Bocquet}, S. and {Brooks}, D. and {Burke}, D.~L. and {Carnero Rosell}, A. and {Carrasco Kind}, M. and {Carretero}, J. and {da Costa}, L.~N. and {Pereira}, M.~E.~S. and {De Vicente}, J. and {Desai}, S. and {Diehl}, H.~T. and {Doel}, P. and {Ferrero}, I. and {Friedel}, D.~N. and {Frieman}, J. and {Garc{\'\i}a-Bellido}, J. and {Gaztanaga}, E. and {Gruen}, D. and {Gutierrez}, G. and {Hinton}, S.~R. and {Hollowood}, D.~L. and {Honscheid}, K. and {James}, D.~J. and {Kuehn}, K. and {Kuropatkin}, N. and {Lidman}, C. and {Maia}, M.~A.~G. and {Menanteau}, F. and {Miquel}, R. and {Morgan}, R. and {Paz-Chinch{\'o}n}, F. and {Pieres}, A. and {Plazas Malag{\'o}n}, A.~A. and {Rodriguez-Monroy}, M. and {Sanchez}, E. and {Scarpine}, V. and {Serrano}, S. and {Sevilla-Noarbe}, I. and {Smith}, M. and {Suchyta}, E. and {Swanson}, M.~E.~C. and {Tarl{\'e}}, G. and {To}, C. and {DES Collaboration}},
        title = "{Correction to: Optical variability of quasars with 20-year photometric light curves}",
      journal = {\mnras},
         year = 2023,
        month = may,
       volume = {521},
       number = {1},
        pages = {836-839},
          doi = {10.1093/mnras/stad592},
       adsurl = {https://ui.adsabs.harvard.edu/abs/2023MNRAS.521..836S}
}

@ARTICLE{Simm2016,
       author = {{Simm}, T. and {Salvato}, M. and {Saglia}, R. and {Ponti}, G. and {Lanzuisi}, G. and {Trakhtenbrot}, B. and {Nandra}, K. and {Bender}, R.},
        title = "{Pan-STARRS1 variability of XMM-COSMOS AGN. II. Physical correlations and power spectrum analysis}",
      journal = {\aap},
         year = 2016,
        month = jan,
       volume = {585},
          eid = {A129},
        pages = {A129},
          doi = {10.1051/0004-6361/201527353},
archivePrefix = {arXiv},
       eprint = {1510.06737},
 primaryClass = {astro-ph.GA},
       adsurl = {https://ui.adsabs.harvard.edu/abs/2016A&A...585A.129S}
}

@ARTICLE{MacLeod2010,
       author = {{MacLeod}, C.~L. and {Ivezi{\'c}}, {\v{Z}}. and {Kochanek}, C.~S. and {Koz{\l}owski}, S. and {Kelly}, B. and {Bullock}, E. and {Kimball}, A. and {Sesar}, B. and {Westman}, D. and {Brooks}, K. and {Gibson}, R. and {Becker}, A.~C. and {de Vries}, W.~H.},
        title = "{Modeling the Time Variability of SDSS Stripe 82 Quasars as a Damped Random Walk}",
      journal = {\apj},
         year = 2010,
        month = oct,
       volume = {721},
       number = {2},
        pages = {1014-1033},
          doi = {10.1088/0004-637X/721/2/1014},
archivePrefix = {arXiv},
       eprint = {1004.0276},
 primaryClass = {astro-ph.CO},
       adsurl = {https://ui.adsabs.harvard.edu/abs/2010ApJ...721.1014M}
}

@ARTICLE{2004ApJ...613..682P,
       author = {{Peterson}, B.~M. and {Ferrarese}, L. and {Gilbert}, K.~M. and {Kaspi}, S. and {Malkan}, M.~A. and {Maoz}, D. and {Merritt}, D. and {Netzer}, H. and {Onken}, C.~A. and {Pogge}, R.~W. and {Vestergaard}, M. and {Wandel}, A.},
        title = "{Central Masses and Broad-Line Region Sizes of Active Galactic Nuclei. II. A Homogeneous Analysis of a Large Reverberation-Mapping Database}",
      journal = {\apj},
         year = 2004,
        month = oct,
       volume = {613},
       number = {2},
        pages = {682-699},
          doi = {10.1086/423269},
archivePrefix = {arXiv},
       eprint = {astro-ph/0407299},
 primaryClass = {astro-ph},
       adsurl = {https://ui.adsabs.harvard.edu/abs/2004ApJ...613..682P}
}

@ARTICLE{Abramowicz1988,
       author = {{Abramowicz}, M.~A. and {Czerny}, B. and {Lasota}, J.~P. and {Szuszkiewicz}, E.},
        title = "{Slim Accretion Disks}",
      journal = {\apj},
         year = 1988,
        month = sep,
       volume = {332},
        pages = {646},
          doi = {10.1086/166683},
       adsurl = {https://ui.adsabs.harvard.edu/abs/1988ApJ...332..646A}
}

@ARTICLE{Du2015,
       author = {{Du}, Pu and {Hu}, Chen and {Lu}, Kai-Xing and {Huang}, Ying-Ke and {Cheng}, Cheng and {Qiu}, Jie and {Li}, Yan-Rong and {Zhang}, Yang-Wei and {Fan}, Xu-Liang and {Bai}, Jin-Ming and {Bian}, Wei-Hao and {Yuan}, Ye-Fei and {Kaspi}, Shai and {Ho}, Luis C. and {Netzer}, Hagai and {Wang}, Jian-Min and {SEAMBH Collaboration}},
        title = "{Supermassive Black Holes with High Accretion Rates in Active Galactic Nuclei. IV. H{\ensuremath{\beta}} Time Lags and Implications for Super-Eddington Accretion}",
      journal = {\apj},
         year = 2015,
        month = jun,
       volume = {806},
       number = {1},
          eid = {22},
        pages = {22},
          doi = {10.1088/0004-637X/806/1/22},
archivePrefix = {arXiv},
       eprint = {1504.01844},
 primaryClass = {astro-ph.GA},
       adsurl = {https://ui.adsabs.harvard.edu/abs/2015ApJ...806...22D}
}

@ARTICLE{Pozo2025,
       author = {{Pozo Nu{\~n}ez}, F. and {Ba{\~n}ados}, E. and {Panda}, S. and {Heidt}, J.},
        title = "{Accretion disc reverberation mapping in a high-redshift quasar}",
      journal = {\aap},
         year = 2025,
        month = aug,
       volume = {700},
          eid = {L8},
        pages = {L8},
          doi = {10.1051/0004-6361/202555421},
archivePrefix = {arXiv},
       eprint = {2508.00209},
 primaryClass = {astro-ph.GA},
       adsurl = {https://ui.adsabs.harvard.edu/abs/2025A&A...700L...8P}
}

@ARTICLE{Gravity2018,
       author = {{Gravity Collaboration} and {Sturm}, E. and {Dexter}, J. and {Pfuhl}, O. and {Stock}, M.~R. and {Davies}, R.~I. and {Lutz}, D. and {Cl{\'e}net}, Y. and {Eckart}, A. and {Eisenhauer}, F. and {Genzel}, R. and {Gratadour}, D. and {H{\"o}nig}, S.~F. and {Kishimoto}, M. and {Lacour}, S. and {Millour}, F. and {Netzer}, H. and {Perrin}, G. and {Peterson}, B.~M. and {Petrucci}, P.~O. and {Rouan}, D. and {Waisberg}, I. and {Woillez}, J. and {Amorim}, A. and {Brandner}, W. and {F{\"o}rster Schreiber}, N.~M. and {Garcia}, P.~J.~V. and {Gillessen}, S. and {Ott}, T. and {Paumard}, T. and {Perraut}, K. and {Scheithauer}, S. and {Straubmeier}, C. and {Tacconi}, L.~J. and {Widmann}, F.},
        title = "{Spatially resolved rotation of the broad-line region of a quasar at sub-parsec scale}",
      journal = {\nat},
         year = 2018,
        month = nov,
       volume = {563},
       number = {7733},
        pages = {657-660},
          doi = {10.1038/s41586-018-0731-9},
archivePrefix = {arXiv},
       eprint = {1811.11195},
 primaryClass = {astro-ph.GA},
       adsurl = {https://ui.adsabs.harvard.edu/abs/2018Natur.563..657G}
}

@ARTICLE{Gravity2024,
       author = {{GRAVITY Collaboration} and {Amorim}, A. and {Bourdarot}, G. and {Brandner}, W. and {Cao}, Y. and {Cl{\'e}net}, Y. and {Davies}, R. and {de Zeeuw}, P.~T. and {Dexter}, J. and {Drescher}, A. and {Eckart}, A. and {Eisenhauer}, F. and {Fabricius}, M. and {Feuchtgruber}, H. and {F{\"o}rster Schreiber}, N.~M. and {Garcia}, P.~J.~V. and {Genzel}, R. and {Gillessen}, S. and {Gratadour}, D. and {H{\"o}nig}, S. and {Kishimoto}, M. and {Lacour}, S. and {Lutz}, D. and {Millour}, F. and {Netzer}, H. and {Ott}, T. and {Paumard}, T. and {Perraut}, K. and {Perrin}, G. and {Peterson}, B.~M. and {Petrucci}, P.~O. and {Pfuhl}, O. and {Prieto}, M.~A. and {Rabien}, S. and {Rouan}, D. and {Santos}, D.~J.~D. and {Shangguan}, J. and {Shimizu}, T. and {Sternberg}, A. and {Straubmeier}, C. and {Sturm}, E. and {Tacconi}, L.~J. and {Tristram}, K.~R.~W. and {Widmann}, F. and {Woillez}, J.},
        title = "{The size-luminosity relation of local active galactic nuclei from interferometric observations of the broad-line region}",
      journal = {\aap},
         year = 2024,
        month = apr,
       volume = {684},
          eid = {A167},
        pages = {A167},
          doi = {10.1051/0004-6361/202348167},
archivePrefix = {arXiv},
       eprint = {2401.07676},
 primaryClass = {astro-ph.GA},
       adsurl = {https://ui.adsabs.harvard.edu/abs/2024A&A...684A.167G}
}

@ARTICLE{Wolf2024,
       author = {{Wolf}, Christian and {Lai}, Samuel and {Onken}, Christopher A. and {Amrutha}, Neelesh and {Bian}, Fuyan and {Hon}, Wei Jeat and {Tisserand}, Patrick and {Webster}, Rachel L.},
        title = "{The accretion of a solar mass per day by a 17-billion solar mass black hole}",
      journal = {Nature Astronomy},
         year = 2024,
        month = apr,
       volume = {8},
        pages = {520-529},
          doi = {10.1038/s41550-024-02195-x},
archivePrefix = {arXiv},
       eprint = {2402.15101},
 primaryClass = {astro-ph.CO},
       adsurl = {https://ui.adsabs.harvard.edu/abs/2024NatAs...8..520W}
}

@ARTICLE{Gravity2026,
       author = {{Gravity+ Collaboration} and {Abd El Dayem}, K. and {Aimar}, N. and {Berdeu}, A. and {Berger}, J.-P. and {Bourdarot}, G. and {Bourget}, P. and {Brandner}, W. and {Cao}, Y. and {Correia}, C. and {Cuevas Cardona}, S. and {Davies}, R. and {Defr{\`e}re}, D. and {Drescher}, A. and {Eckart}, A. and {Eisenhauer}, F. and {Fabricius}, M. and {Farah}, A. and {Feuchtgruber}, H. and {F{\"o}rster Schreiber}, N.~M. and {Foschi}, A. and {Garcia}, P. and {Garcia Lopez}, R. and {Genzel}, R. and {Gillessen}, S. and {Gomes}, T. and {Gont{\'e}}, F. and {Gopinath}, V. and {Graf}, J. and {Hartl}, M. and {Haubois}, X. and {Hau{\ss}mann}, F. and {Ho}, L.~C. and {H{\"o}nig}, S. and {Houll{\'e}}, M. and {Joharle}, S. and {Keiman}, C. and {Kervella}, P. and {Kolb}, J. and {Kreidberg}, L. and {Labdon}, A. and {Lacour}, S. and {Lai}, O. and {Lai}, S. and {Laugier}, R. and {Le Bouquin}, J.-B. and {Leftley}, J. and {Li}, R. and {Lopez}, B. and {Lutz}, D. and {Mang}, F. and {M{\'e}rand}, A. and {Millour}, F. and {Montarg{\`e}s}, M. and {More}, N. and {Moruj{\~a}o}, N. and {Nowacki}, H. and {Nowak}, M. and {Oberti}, S. and {Onken}, C. and {Osorno}, J. and {Ott}, T. and {Paumard}, T. and {Perraut}, K. and {Perrin}, G. and {Petrov}, R. and {Petrucci}, P.-O. and {Pourr{\'e}}, N. and {Rabien}, S. and {Rau}, C. and {Ribeiro}, D.~C. and {Robbe-Dubois}, S. and {Sadun Bordoni}, M. and {Salman}, M. and {Sanchez-Bermudez}, J. and {Santos}, D. and {Sauter}, J. and {Scialpi}, M. and {Scigliuto}, J. and {Shangguan}, J. and {Shchekaturov}, P. and {Shimizu}, T. and {Soulez}, F. and {Straubmeier}, C. and {Sturm}, E. and {Subroweit}, M. and {Sykes}, C. and {Tacconi}, L.~J. and {{\"U}bler}, H. and {Ulbricht}, G. and {Vincent}, F. and {Webster}, R. and {Wieprecht}, E. and {Woillez}, J. and {Wolf}, C.},
        title = "{Spatially resolved broad-line region in a quasar at z = 4: Dynamical black hole mass and prominent outflow}",
      journal = {\aap},
         year = 2026,
        month = feb,
       volume = {706},
          eid = {A99},
        pages = {A99},
          doi = {10.1051/0004-6361/202557285},
archivePrefix = {arXiv},
       eprint = {2509.13911},
 primaryClass = {astro-ph.GA},
       adsurl = {https://ui.adsabs.harvard.edu/abs/2026A&A...706A..99G}
}

@ARTICLE{Abuter2024,
       author = {{Abuter}, R. and {Allouche}, F. and {Amorim}, A. and {Bailet}, C. and {Berdeu}, A. and {Berger}, J.-P. and {Berio}, P. and {Bigioli}, A. and {Boebion}, O. and {Bolzer}, M.-L. and {Bonnet}, H. and {Bourdarot}, G. and {Bourget}, P. and {Brandner}, W. and {Cao}, Y. and {Conzelmann}, R. and {Comin}, M. and {Cl{\'e}net}, Y. and {Courtney-Barrer}, B. and {Davies}, R. and {Defr{\`e}re}, D. and {Delboulb{\'e}}, A. and {Delplancke-Str{\"o}bele}, F. and {Dembet}, R. and {Dexter}, J. and {de Zeeuw}, P.~T. and {Drescher}, A. and {Eckart}, A. and {{\'E}douard}, C. and {Eisenhauer}, F. and {Fabricius}, M. and {Feuchtgruber}, H. and {Finger}, G. and {F{\"o}rster Schreiber}, N.~M. and {Garcia}, P. and {Garcia Lopez}, R. and {Gao}, F. and {Gendron}, E. and {Genzel}, R. and {Gil}, J.~P. and {Gillessen}, S. and {Gomes}, T. and {Gont{\'e}}, F. and {Gouvret}, C. and {Guajardo}, P. and {Guieu}, S. and {Hackenberg}, W. and {Haddad}, N. and {Hartl}, M. and {Haubois}, X. and {Hau{\ss}mann}, F. and {Hei{\ss}el}, G. and {Henning}, Th. and {Hippler}, S. and {H{\"o}nig}, S.~F. and {Horrobin}, M. and {Hubin}, N. and {Jacqmart}, E. and {Jocou}, L. and {Kaufer}, A. and {Kervella}, P. and {Kolb}, J. and {Korhonen}, H. and {Lacour}, S. and {Lagarde}, S. and {Lai}, O. and {Lapeyr{\`e}re}, V. and {Laugier}, R. and {Le Bouquin}, J.-B. and {Leftley}, J. and {L{\'e}na}, P. and {Lewis}, S. and {Liu}, D. and {Lopez}, B. and {Lutz}, D. and {Magnard}, Y. and {Mang}, F. and {Marcotto}, A. and {Maurel}, D. and {M{\'e}rand}, A. and {Millour}, F. and {More}, N. and {Netzer}, H. and {Nowacki}, H. and {Nowak}, M. and {Oberti}, S. and {Ott}, T. and {Pallanca}, L. and {Paumard}, T. and {Perraut}, K. and {Perrin}, G. and {Petrov}, R. and {Pfuhl}, O. and {Pourr{\'e}}, N. and {Rabien}, S. and {Rau}, C. and {Riquelme}, M. and {Robbe-Dubois}, S. and {Rochat}, S. and {Salman}, M. and {Sanchez-Bermudez}, J. and {Santos}, D.~J.~D. and {Scheithauer}, S. and {Sch{\"o}ller}, M. and {Schubert}, J. and {Schuhler}, N. and {Shangguan}, J. and {Shchekaturov}, P. and {Shimizu}, T.~T. and {Sevin}, A. and {Soulez}, F. and {Spang}, A. and {Stadler}, E. and {Sternberg}, A. and {Straubmeier}, C. and {Sturm}, E. and {Sykes}, C. and {Tacconi}, L.~J. and {Tristram}, K.~R.~W. and {Vincent}, F. and {von Fellenberg}, S. and {Uysal}, S. and {Widmann}, F. and {Wieprecht}, E. and {Wiezorrek}, E. and {Woillez}, J. and {Zins}, G.},
        title = "{A dynamical measure of the black hole mass in a quasar 11 billion years ago}",
      journal = {\nat},
         year = 2024,
        month = mar,
       volume = {627},
       number = {8003},
        pages = {281-285},
          doi = {10.1038/s41586-024-07053-4},
archivePrefix = {arXiv},
       eprint = {2401.14567},
 primaryClass = {astro-ph.GA},
       adsurl = {https://ui.adsabs.harvard.edu/abs/2024Natur.627..281A}
}

@ARTICLE{Bai2026,
       author = {{Bai}, Hua-Rui and {Du}, Pu and {Hu}, Chen and {Chen}, Yong-Jie and {Yao}, Zhu-Heng and {Li}, Yan-Rong and {Fu}, Yi-Xin and {Wang}, Yi-Lin and {Zhao}, Yu and {Zhang}, Hao and {Liu}, Jun-Rong and {Yang}, Sen and {Peng}, Yue-Chang and {Fang}, Feng-Na and {Songsheng}, Yu-Yang and {Xiao}, Ming and {Zhai}, Shuo and {Li}, Sha-Sha and {Lu}, Kai-Xing and {Zhang}, Zhi-Xiang and {Bao}, Dong-Wei and {Guo}, Wei-Jian and {Feng}, Jia-Qi and {Zhao}, Yi-Peng and {Aceituno}, Jes{\'u}s and {Bai}, Jin-Ming and {Ho}, Luis C. and {Wang}, Jian-Min and {Seambh Collaboration}},
        title = "{Supermassive Black Holes with High Accretion Rates in Active Galactic Nuclei. XV. Reverberation Mapping of Mg II Emission Lines}",
      journal = {\apjs},
         year = 2026,
        month = feb,
       volume = {282},
       number = {2},
          eid = {56},
        pages = {56},
          doi = {10.3847/1538-4365/ae2b6d},
archivePrefix = {arXiv},
       eprint = {2512.08192},
 primaryClass = {astro-ph.GA},
       adsurl = {https://ui.adsabs.harvard.edu/abs/2026ApJS..282...56B}
}

@ARTICLE{Sadowski2009,
       author = {{S{\k{a}}dowski}, Aleksander},
        title = "{Slim Disks Around Kerr Black Holes Revisited}",
      journal = {\apjs},
         year = 2009,
        month = aug,
       volume = {183},
       number = {2},
        pages = {171-178},
          doi = {10.1088/0067-0049/183/2/171},
archivePrefix = {arXiv},
       eprint = {0906.0355},
 primaryClass = {astro-ph.HE},
       adsurl = {https://ui.adsabs.harvard.edu/abs/2009ApJS..183..171S}
}

@ARTICLE{Pozo2019,
       author = {{Pozo Nu{\~n}ez}, F. and {Gianniotis}, N. and {Blex}, J. and {Lisow}, T. and {Chini}, R. and {Polsterer}, K.~L. and {Pott}, J. -U. and {Esser}, J. and {Pietrzy{\'n}ski}, G.},
        title = "{Optical continuum photometric reverberation mapping of the Seyfert-1 galaxy Mrk509}",
      journal = {\mnras},
         year = 2019,
        month = dec,
       volume = {490},
       number = {3},
        pages = {3936-3951},
          doi = {10.1093/mnras/stz2830},
archivePrefix = {arXiv},
       eprint = {1912.10319},
 primaryClass = {astro-ph.GA},
       adsurl = {https://ui.adsabs.harvard.edu/abs/2019MNRAS.490.3936P}
}

@ARTICLE{Chelouche2019,
       author = {{Chelouche}, Doron and {Pozo Nu{\~n}ez}, Francisco and {Kaspi}, Shai},
        title = "{Direct evidence of non-disk optical continuum emission around an active black hole}",
      journal = {Nature Astronomy},
         year = 2019,
        month = jan,
       volume = {3},
        pages = {251-257},
          doi = {10.1038/s41550-018-0659-x},
       adsurl = {https://ui.adsabs.harvard.edu/abs/2019NatAs...3..251C}
}

@ARTICLE{Gaskell1987,
       author = {{Gaskell}, C. Martin and {Peterson}, Bradley M.},
        title = "{The Accuracy of Cross-Correlation Estimates of Quasar Emission-Line Region Sizes}",
      journal = {\apjs},
         year = 1987,
        month = sep,
       volume = {65},
        pages = {1},
          doi = {10.1086/191216},
       adsurl = {https://ui.adsabs.harvard.edu/abs/1987ApJS...65....1G}
}

@ARTICLE{Korista2019,
       author = {{Korista}, K.~T. and {Goad}, M.~R.},
        title = "{Quantifying the impact of variable BLR diffuse continuum contributions on measured continuum interband delays}",
      journal = {\mnras},
         year = 2019,
        month = nov,
       volume = {489},
       number = {4},
        pages = {5284-5300},
          doi = {10.1093/mnras/stz2330},
archivePrefix = {arXiv},
       eprint = {1908.07757},
 primaryClass = {astro-ph.GA},
       adsurl = {https://ui.adsabs.harvard.edu/abs/2019MNRAS.489.5284K}
}

@INPROCEEDINGS{1997ASSL..218..163A,
       author = {{Alexander}, Tal},
        title = "{Is AGN Variability Correlated with Other AGN Properties? ZDCF Analysis of Small Samples of Sparse Light Curves}",
    booktitle = {Astronomical Time Series},
         year = 1997,
       editor = {{Maoz}, D. and {Sternberg}, A. and {Leibowitz}, E.~M.},
       series = {Astrophysics and Space Science Library},
       volume = {218},
        month = jan,
        pages = {163},
          doi = {10.1007/978-94-015-8941-3_14},
       adsurl = {https://ui.adsabs.harvard.edu/abs/1997ASSL..218..163A}
}

@ARTICLE{2017ApJ...844..146C,
       author = {{Chelouche}, Doron and {Pozo-Nu{\~n}ez}, Francisco and {Zucker}, Shay},
        title = "{Methods of Reverberation Mapping. I. Time-lag Determination by Measures of Randomness}",
      journal = {\apj},
         year = 2017,
        month = aug,
       volume = {844},
       number = {2},
          eid = {146},
        pages = {146},
          doi = {10.3847/1538-4357/aa7b86},
archivePrefix = {arXiv},
       eprint = {1708.04477},
 primaryClass = {astro-ph.IM},
       adsurl = {https://ui.adsabs.harvard.edu/abs/2017ApJ...844..146C}
}

\end{document}